%% file: main.tex
\documentclass{article}
\usepackage[top=1in, bottom=1in, left=1in, right=1in]{geometry}
\usepackage{tgpagella}
\usepackage[utf8]{inputenc} 
\usepackage[T1]{fontenc}    
\usepackage{bm}
\usepackage{pdfsync}
\input{macro}

\title{Statistical and Algorithmic Foundations of Probing Quantum Systems with Compressive Measurements: A Review}

\author{Zhen Qin, Michael B. Wakin, and Zhihui Zhu\thanks{ZQ (email: zhenqin@umich.edu) is with the Michigan Institute for Computational Discovery and Engineering, Department of Electrical Engineering and Computer Science, Department of Statistics, University of Michigan, Ann Arbor, MI 48109 USA; MBW (email: mwakin@mines.edu) is with the Department of Electrical Engineering, Colorado School of Mines; and ZZ (email: zhu.3440@osu.edu) is with the Department of Computer Science and Engineering, the Ohio State University.}
}

\begin{document}

\maketitle

\begin{abstract}
Quantum state tomography (QST) is a fundamental task in quantum information science that aims to reconstruct unknown quantum states from measurement data. However, the exponential growth of Hilbert-space dimension with system size makes full tomography of general quantum states statistically and computationally prohibitive. This challenge has motivated extensive research on structured quantum state tomography, where prior structure, such as low-rankness, tensor-network representations, shallow quantum circuits, and neural quantum states, can substantially reduce the effective degrees of freedom and enable scalable recovery. In this review, we provide a unified perspective on QST for structured quantum states through three closely related themes: compact state representations, measurement design, and computational algorithms. After reviewing common models for structured quantum states, we survey existing work on geometric preservation properties of measurement frameworks, ranging from informationally complete POVMs to randomized measurements, and their implications for sample complexity. On the algorithmic side, we review optimization methods for reconstructing structured quantum states from empirical measurements. By connecting QST with broader principles from compressive sensing, matrix sensing, and structured inverse problems, this survey highlights common theoretical foundations underlying sample complexity, measurement efficiency, and scalable recovery.
\end{abstract}

\tableofcontents

\section{Introduction}
Quantum information science leverages the principles of quantum mechanics to harness the inherent advantages of quantum systems \cite{nielsen2010quantum}. This rapidly evolving field has seen remarkable progress in recent years, with advances in quantum computing \cite{ladd2010quantum}, quantum communication \cite{gisin2007quantum}, and quantum sensing technologies \cite{degen2017quantum}. For example, driven by advances in hardware and experimental techniques, the size of quantum computers has rapidly increased in recent years, with some of the most advanced processors having over 100 qubits \cite{preskill2018quantum,chow2021ibm}. On one hand, the emergence of these advanced programmable quantum simulators and computers \cite{bloch2012quantum} offers unprecedented potential in exploring and harnessing the intricacies of highly entangled complex quantum systems. In the foreseeable future, the advent of large-scale error-corrected quantum computers \cite{preskill2018quantum} will propel us forward in exploring the frontiers of quantum physical science and practical applications. On the other hand, the inherent complexity also poses challenges in precise control and accurate characterization of these systems. Leveraging tools from multidisciplinary fields such as physics, computer science, and applied mathematics becomes paramount in the development of techniques to efficiently probe the properties of complex quantum systems.

Measuring and characterizing quantum systems presents a long-standing challenge in quantum information science \cite{james2001measurement}. The reconstruction of quantum states from experimental measurements, often achieved using quantum state tomography (QST), remains a useful subroutine for building, calibrating, benchmarking and controlling quantum information processing devices. The goal of QST is to find a complete density matrix (a classical description for a system), from which all observable properties can be extracted via classical calculations \cite{bertrand1987tomographic,vogel1989determination,leonhardt1995quantum,hradil1997quantum,james2001measurement}. However, the exponential scaling of the Hilbert space dimension with the number of qubits makes full characterization extremely challenging; for an $n$-qudit quantum system, the density matrix lies in $\C^{d^n\times d^n}$. This challenge is further compounded by the probabilistic nature of quantum mechanics: information about a quantum system can only be accessed through repeated measurements, each of which produces a stochastic outcome governed by the underlying quantum state. Consequently, direct tomography of arbitrary quantum states typically incurs exponential sample, storage, and computational complexity, making conventional approaches intractable for large-scale systems. This challenge motivates a central question: \emph{how can one exploit structures in physically relevant quantum states to enable scalable tomography?}

A key observation is that many physically relevant quantum states exhibit low-complexity structure and admit compact representations. Examples include low-rank states associated with low-entropy or nearly pure systems \cite{flammia2012quantum}, tensor-network representations such as matrix product states (MPS) \cite{perez2007matrix}, matrix product operators (MPOs) \cite{pirvu2010matrix}, and projected entangled-pair operators (PEPOs) \cite{cirac2021matrix}, states generated by shallow quantum circuits \cite{huang2024learning}, and more recently neural quantum states represented by neural networks \cite{lange2024architectures}. These representations often reduce the intrinsic degrees of freedom far below the ambient dimension of the Hilbert space, suggesting that the complexity of tomography could be governed by the effective structure of the state rather than by the full dimension of the density matrix.

Meanwhile, efficient tomography depends not only on the representation of the target state, but also on measurement design and reconstruction algorithms. Different measurement settings, including informationally complete POVMs \cite{matthews2009distinguishability,kueng2017low}, Pauli measurements \cite{guctua2020fast}, random unitary measurements \cite{gross2007evenly,dankert2009exact,roy2009unitary}, Haar-random projective measurements \cite{voroninski2013quantum}, local measurements \cite{lancien2013distinguishing,brandao2020fast,jameson2024optimal}, and adaptive measurement strategies \cite{chen2023does,chen2024adaptivity}, lead to fundamentally different trade-offs between experimental feasibility, statistical efficiency, and computational complexity. Likewise, effectively exploiting low-dimensional structure for tomography requires the development of efficient optimization and reconstruction methods. Understanding the interplay among \emph{state structure, measurement design, sample complexity, and optimization} is therefore central to scalable QST.

While QST possesses several unique features, such as probabilistic quantum measurements and physical constraints that will be described in detail in \Cref{sec:experiment-setting}, it can also be viewed more broadly as a structured inverse problem. From this perspective, QST shares deep conceptual and mathematical connections with inverse problems studied in signal processing, machine learning, and applied mathematics, particularly compressive sensing \cite{baraniuk2007compressive} and low-rank matrix recovery \cite{park2017non,chi2019nonconvex}. Similar to QST, three tightly coupled components---\emph{structure}, \emph{measurement design}, and \emph{algorithm design}---play a central role in modern inverse problems. First, structure reduces the intrinsic complexity of the target signal by restricting attention to a low-dimensional or low-complexity model class, such as sparsity, low-rankness, manifold constraints, or other compact parameterizations. As a result, the statistical complexity of recovery often scales with the effective degrees of freedom rather than the ambient dimension \cite{donoho2006compressed,candes2006near,candes2011tight,boumal2023introduction}. Geometric complexity measures such as covering numbers, intrinsic dimension, and manifold complexity provide principled ways to quantify this effective complexity. Second, measurement design determines whether the measurement operator preserves sufficient information about the structured signal. In particular, embedding properties such as restricted isometry properties (RIP) \cite{candes2008restricted}, stable embeddings \cite{tropp2015convex,eftekhari2015new}, and related geometric preservation conditions \cite{matthews2009distinguishability,baraniuk2009random,kueng2017low,huang2020predicting} provide a unified framework for understanding when structured signals remain distinguishable after dimensionality reduction or indirect observation \cite{baraniuk2008simple}. Third, algorithm design translates these statistical and geometric properties into practical recovery procedures. A broad range of methods, including convex relaxations \cite{chen2001atomic,fazel2002matrix,candes2006robust,recht2010guaranteed}, projected methods \cite{aharoni1989block,lewis2008alternating,foucart2011hard}, iterative thresholding \cite{blumensath2009iterative,cai2010singular,rauhut2017low}, and factorized nonconvex optimization \cite{park2018finding,haeffele2019structured,han2022optimal,qin2025scalable}, have been developed to efficiently reconstruct structured signals from incomplete or noisy measurements by explicitly exploiting low-dimensional structure.

These general principles of structured inverse problems have also shaped recent developments in QST. On the representation side, new parameterized models have been developed to capture increasingly complex quantum-state structures, particularly those that exploit the expressive power of neural networks \cite{lange2024architectures}. On the measurement side, randomized measurement ensembles have been widely adopted in recent years \cite{elben2023randomized}, while tools such as restricted isometry properties (RIP), stable embeddings, and information-geometric characterizations have provided increasingly unified principles for analyzing when measurement ensembles preserve sufficient information for stable recovery \cite{qin2024quantum,qin2024sample,qin2025quantum,qin2026optimal}. On the algorithmic side, projected methods \cite{sugiyama2013precision,guctua2020fast}, gradient-based optimization \cite{cramer2010efficient,qin2024sample}, and classical-shadow-based estimators \cite{huang2020predicting,qin2025enhancing} have enabled scalable reconstruction procedures with increasingly strong theoretical guarantees.

This review aims to provide a unified perspective on QST for structured quantum states through three closely related themes: \emph{compact representation}, \emph{measurement design}, and \emph{computational algorithms}. These three components are tightly coupled: compact state representations reduce intrinsic statistical complexity; informative measurement ensembles preserve this structure in the observed data; and efficient reconstruction algorithms translate these properties into practical recovery guarantees.  Compared with prior surveys that focus more broadly on quantum certification, benchmarking, or general tomography methodologies \cite{eisert2020quantum,anshu2024survey}, our review emphasizes a unified inverse-problem perspective on structured QST, highlighting the interplay among compact state representations, measurement geometry, sample complexity, and optimization-based recovery. Beyond summarizing recent theoretical developments, this survey is intended to bridge communities across quantum information, signal processing, and machine learning. Thus, we provide a linear-algebraic and optimization-oriented description of key quantum mechanisms underlying QST, including quantum states, density-matrix representations, measurement operators, POVMs, and randomized measurement ensembles. By presenting these concepts from the perspective of inverse problems and structured recovery, we aim to make recent developments in QST accessible to a broader research community beyond quantum physics. More generally, we believe QST provides a rich setting in which statistical complexity, geometric measurement design, physical constraints, and computational tractability interact in a highly nontrivial way.

The rest of this review is organized as follows. In \Cref{sec:experiment-setting}, we first introduce the basic quantum-mechanical concepts needed to understand QST and provide a detailed description of practical measurement settings. In \Cref{sec: quantum state tomography}, we summarize common structured quantum-state representations, review how structure influences sample complexity and measurement design, and discuss optimization methods for reconstruction. Finally, we conclude in \Cref{sec:discussion-conclusion} with open problems and future directions, including local measurement analysis, convergence theory for complex structured models, neural quantum representations, and adaptive measurement design.

\section{Experimental Recipe for Characterizing Quantum Systems}\label{sec:experiment-setting}
We first provide an overview of the protocol for acquiring quantum measurements.

\subsection{Protocol---Measurement acquisition}

\paragraph{States and density operators}
In quantum physics, an isolated quantum system is described by a state vector $\ket{\psi}$, represented in Dirac notation. This vector corresponds to a unit-length element in a complex vector space, commonly referred to as the Hilbert space. As an example, a \emph{qubit}, the simplest quantum system, is associated with a vector in a two-dimensional Hilbert space. The basis vectors of this space are often chosen as $\ket{0}$ and $\ket{1}$, which represent two distinct physical states of the qubit, such as the ground state and the first excited state of an atom. Any qubit state can be written as $\ket{\psi} = a\ket{0} + b\ket{1}$, where the coefficients $a$ and $b$ are complex numbers that satisfy the normalization condition $|a|^2 + |b|^2 = 1$. Equivalently, the state vector $\ket{\psi}$ can be expressed as a $2 \times 1$ column vector
$\vpsi := \begin{bmatrix}a\\ b\end{bmatrix} \in \C^2.$

A \emph{qudit} generalizes the concept of a qubit to a $d$-dimensional system, represented in a $d$-dimensional Hilbert space. Each state vector in this space is equivalently expressed as a unit-length vector in $\C^d$. While qubits are the standard information units in most quantum computers, akin to bits in classical computation, this paper adopts qudits to allow for a more general framework. For quantum many-body systems, the total state space is represented as the tensor product of the individual qudit state spaces. Specifically, in a composite system of $\nqbit$ qudits, the state vector $\vpsi$ resides in $\C^{d^\nqbit}$ and satisfies the unit-length condition.

Until now we have considered quantum systems that are fully described by a state vector $\vpsi$, corresponding to a \emph{pure state}. However, quantum systems may also exist in \emph{mixed states}, where the system occupies one of several states $\vpsi_i$ with associated probabilities $\alpha_i$. Such a mixed state is characterized by the ensemble $\{\alpha_i, \vpsi_i  \}$, where $0 \leq \alpha_i \leq 1$ and $\sum_i \alpha_i = 1$. A quantum system in a mixed state is described by a \emph{density operator} or \emph{density matrix}.\footnote{Formally, a density matrix represents a density operator with respect to a chosen basis in the underlying Hilbert space. Throughout this paper, we adopt the standard computational basis for qudits, denoted as $\{|0\rangle, |1\rangle, \dots, |d-1\rangle\}$. Consequently, the terms \emph{density matrix} and \emph{density operator} are used interchangeably.}  The density operator of a pure state $\vpsi\in \C^{d^\nqbit}$ is given by
\begin{eqnarray}
\label{pure state density matrix}
\vrho = \vpsi \vpsi^\dagger \in \C^{d^\nqbit \times d^\nqbit}.
\end{eqnarray}
For a mixed state, the density operator can be written as
\begin{eqnarray}
\label{mixed state density matrix}
\vrho = \sum_i \alpha_i \vpsi_i \vpsi_i^\dagger \in \C^{d^\nqbit \times d^\nqbit}.
\end{eqnarray}
A density operator with rank one corresponds to a pure state, while one with rank greater than one represents a mixed state. In all cases, we have that $(i)$~the density operator $\vrho \succeq \vzero$ is a positive semidefinite (PSD) matrix, and $(ii)$~$\trace(\vrho) = 1$.

\paragraph{Quantum measurements}

To determine $\vrho$ or characterize the properties of an experimental quantum state, it is generally necessary to perform quantum measurements on a large number of identical copies of the state. This necessity arises because quantum measurements are invasive and modify the state being measured, much like polarized light passing through a polarizing filter is irreversibly altered. The most general measurements one can physically perform on a quantum system are described by Positive Operator Valued Measures (POVMs) \cite{nielsen2002quantum}, as explained below.

\begin{defi} [POVM \cite{nielsen2002quantum}] A Positive Operator Valued Measure (POVM) is a set of PSD matrices $\{\mA_1,\ldots,\mA_K \}$ such that
\begin{eqnarray}
\label{The defi of POVM 1}
\sum_{k=1}^K \mA_k = \mId.
\end{eqnarray}
Each POVM element $\mA_k$ is associated with a possible outcome of a quantum measurement, and the probability $p_k$ of detecting the $k$-th outcome when measuring the density operator $\vrho$ is given by
\begin{eqnarray}
\label{The defi of POVM 2}
p_k = \innerprod{\mA_k}{\vrho},
\end{eqnarray}
where $\sum_{k=1}^Kp_k=1$ due to \eqref{The defi of POVM 1} and the fact that $\trace(\vrho) = 1$. We often repeat the measurement process $M$ times and take the average of the statistically independent outcomes to generate the empirical probabilities
\begin{equation}
\wh p_{k} = \frac{f_k}{M}, \  k \in[K]:=\{1,\ldots,K\},
\label{eq:empirical-prob}\end{equation}
where $f_k$ denotes the number of times the $k$-th outcome is observed. In information theory and signal processing communities, one might call $\{p_k\}$ and $\{\wh p_k\}$ the population and empirical (linear) measurements, respectively.
\label{def:POVM}\end{defi}

Collectively, the random variables $f_1,\ldots,f_K$ are characterized by the multinomial distribution $\operatorname{Multinom}$ $\operatorname{-ial}(M, \vp)$ \cite{severini2005elements} with parameters $M$ and $\vp = \begin{bmatrix} p_1 & \cdots & p_K\end{bmatrix}^\top$, where $p_k$ is defined in \eqref{The defi of POVM 2}. Without considering the statistical error, $\{p_k\}$ can be viewed as $K$ linear measurements of the state~$\vrho$. 

If a POVM is informationally complete, it alone suffices to recover any quantum state as long as a sufficiently large number of state copies are taken. However, as will be described later in details about the choices of POVMs, in many other cases, an individual POVM, such as projective rank-one measurements, might be insufficient to recover a general quantum state $\vrho$ even if we repeat the measurement infinitely many times, since the probabilities $\{p_k\}$ only provide us the diagonal elements of $\vrho$ in the basis formed by the standard computational basis for the qudits, denoted by $\{|0\rangle, |1\rangle, \cdots, |d-1\rangle\}$. Therefore measuring with multiple POVMs can be used to acquire a more holistic understanding of the quantum state.

For simplicity, consider $Q$ POVMs, indexed by $q\in[Q]$, where each POVM $\{\mA_{q,k}\}_{k\in[K]}$ contains the same number of PSD operators $K$ and is probed with $M$ shots, returning the empirical frequencies $\wh\vp_q$ for $q \in [Q]$ where the bold notation indicates a vector: $\wh \vp_q = [\wh p_{q,1} \cdots \wh p_{q,K}]^\top$. To simplify the notation, we collect the probabilities for each POVM $\{ \<\mA_{q,k}, \vrho\>\}$, into a single linear map $\calA_q:  \C^{d^n\times  d^n} \rightarrow \R^K$ of the form
\begin{eqnarray}
\label{The defi of POVM element in q-th POVM}
 \calA_q(\vrho) =  \begin{bmatrix}
          \< \mA_{q,1}, \vrho  \> \\
          \vdots \\
          \< \mA_{q,K}, \vrho  \>
        \end{bmatrix}.
\end{eqnarray}
Stacking the linear operators $\{\calA_q\}$ corresponding to the $Q$ POVMs as a single linear map  $ \calA:\C^{d^n\times d^n} \rightarrow \R^{KQ}$, we can generate $K Q$ population measurements as
\begin{eqnarray}
\label{The defi of population measurement in Q cases (K measurements)}
 \vp = \calA(\vrho)  = \begin{bmatrix}
          {\bm p}_{1} \\
          \vdots \\
          {\bm p}_{Q}
        \end{bmatrix}  = \begin{bmatrix}
          \calA_1(\vrho) \\
          \vdots \\
          \calA_Q(\vrho)
        \end{bmatrix}.
\end{eqnarray}
For each POVM, we repeat the measurement process $M$ times and stack  all the total empirical measurements together as\footnote{To simplify notation, we assume each POVM contains $K$ PSD operators and is performed $M$ times; however, both $M$ and $K$ may vary between different POVMs.} 
\begin{equation}
\wh\vp = \begin{bmatrix}
    \wh \vp_1 \\ \vdots \\ \wh \vp_Q
\end{bmatrix}.
\label{eq:map-M-POVM1}\end{equation}

In this review, the objective of quantum learning is to estimate a quantum state $\vrho$ by utilizing empirical measurements $\wh \vp$ obtained at a limited experimental cost and processed on a classical computer.

\paragraph{Example: Polarizing filters}

The interactions of polarized light with polarizing filters provide an excellent interpretable case study for concepts involving quantum state vectors, density operators, and quantum measurements.

Any linearly polarized light (horizontal, vertical, or any specific angle) is in a pure state. It can be described by a $2 \times 1$ state vector of the form:
\begin{eqnarray}
\label{polarizedvec}
\vpsi := \begin{bmatrix} \cos(\theta) \\ \sin(\theta) \end{bmatrix} \in \C^2,
\end{eqnarray}
where $\theta$ represents the angle of the polarization relative to the horizontal direction. Equivalently, this $2 \times 1$ vector encodes the coefficients of $\ket{\psi} = \cos(\theta)\ket{0} + \sin(\theta)\ket{1}$ in the orthonormal basis formed by $\ket{0}$ and $\ket{1}$, which correspond to horizontal and vertical linearly polarized light, respectively. Note that the state is normalized, since $\|\vpsi\|_2 = 1$ for any value of $\theta$.

When linearly polarized light arrives at a linearly polarized filter, each photon is either transmitted or absorbed. Let $\alpha$ represent the polarization angle of the filter. Define two state vectors representing the polarization direction of the filter and its orthogonal (blocked) direction:
\begin{eqnarray*}
    \vphi =
    \begin{bmatrix}
    \cos(\alpha) \\
    \sin(\alpha)
    \end{bmatrix} \qquad \text{and} \qquad
    \vphi_\perp =
    \begin{bmatrix}
    \cos(\alpha + \pi/2) \\
    \sin(\alpha + \pi/2)
    \end{bmatrix}
    =
    \begin{bmatrix}
    -\sin(\alpha) \\
    \cos(\alpha)
    \end{bmatrix},
\end{eqnarray*}
and note that $\langle \vphi, \vphi_\perp \rangle = 0$. The probability that each photon is transmitted through the filter is then given by the squared magnitude of the inner product between the two state vectors,
    \begin{eqnarray}
    P_{\text{transmit}} = |\langle \vphi, \vpsi \rangle |^2 = \cos^2(\theta - \alpha),
    \label{eq:malus}
    \end{eqnarray}
and the probability that the photon is absorbed by the filter is given by
    \begin{eqnarray}
    P_{\text{absorb}} = |\langle \vphi_\perp, \vpsi \rangle |^2 = \sin^2(\theta - \alpha) = 1 - P_{\text{transmit}}.
    \label{eq:malus2}
    \end{eqnarray}
This expression in~\eqref{eq:malus} is known as Malus’s Law, which states that the transmitted intensity of linearly polarized light through a polarizer varies as the square of the cosine of the angle between their polarization directions.

We can treat the interaction of each photon with the polarizing filter as a quantum measurement in the following sense. Define the $2 \times 2$ density matrix $\vrho = \vpsi \vpsi^\dagger$, and let $\mA_1 = \vphi \vphi^\dagger$ and $\mA_2 = \vphi_\perp \vphi_\perp^\dagger$. Together, $\mA_1$ and $\mA_2$ form a POVM, since $\mA_1 + \mA_2 = \mId$. One can then verify that
\begin{equation}
P_{\text{transmit}} = \langle \mA_1, \vrho \rangle
\qquad\text{and}\qquad
P_{\text{absorb}} = \langle \mA_2, \vrho \rangle.
\label{eq:phtxab}
\end{equation}
Moreover, after a photon is either transmitted or absorbed, its post-measurement state vector collapses onto either $\vphi$ or $\vphi_\perp$; all information about the original polarization angle $\theta$ is lost. This illustrates why multiple identical copies of a quantum state are required to reconstruct it through quantum state tomography.

While pure states suffice to describe linearly polarized light, mixed states are required to describe partially polarized or unpolarized light. For example, light containing $50\%$ horizontally polarized photons (denoted with state vector $\vpsi_{\text{H}} = \begin{bmatrix} 1 \\ 0 \end{bmatrix}$) and $50\%$ vertically polarized photons (denoted with state vector $\vpsi_{\text{V}} = \begin{bmatrix} 0 \\ 1 \end{bmatrix}$) can be described by a density operator
\begin{equation}
\vrho = \frac{1}{2} \vpsi_{\text{H}} \vpsi_{\text{H}}^\dagger + \frac{1}{2} \vpsi_{\text{V}} \vpsi_{\text{V}}^\dagger = \frac{1}{2} \begin{bmatrix} 1 & 0 \\ 0 & 1 \end{bmatrix}.
\label{eq:rhomixedph}
\end{equation}
Because $\vrho$ has rank $2$, it cannot be factored into an outer product of any vector with itself. Such a quantum system is said to be in a mixed (as opposed to pure) state. While~\eqref{eq:malus} and~\eqref{eq:malus2} can no longer be directly applied to calculate the transmission and absorption probabilities, the equations in~\eqref{eq:phtxab} carry over exactly. Indeed, both $P_{\text{transmit}}$ and $P_{\text{absorb}}$ equal $1/2$ in this scenario, and due to the decomposition of $\vrho$ in~\eqref{eq:rhomixedph}, one can decompose each probability in~\eqref{eq:phtxab} similarly. For example, $P_{\text{transmit}}$ equals $1/2$ times the probability of a horizontally polarized photon being transmitted plus $1/2$ times the probability of a vertically polarized photon being transmitted. Such light is said to be unpolarized, since no polarization direction is preferred. Moreover, the same density matrix (and thus for all practical purposes, the same light) can be constructed through an equal mixture of any two orthogonal directions of linearly polarized light: not only $\text{H}$ and $\text{V}$, but any $\theta$ and $\theta + \pi/2$. Nonequal mixtures of polarization states lead to partially polarized light.

\paragraph{Comparison with linear measurements in signal processing}

To recap, the QST problem involves estimating a quantum state $\vrho$ based on empirical measurements $\wh \vp$. The empirical measurements are drawn randomly from a multinomial probability distribution controlled by the population measurements $\vp = \calA(\vrho)$. Moreover, $\mathbb{E}[\wh \vp] = \vp$. For a fixed measurement setting $\calA$, the population measurements are not random, and they depend linearly on the unknown state $\vrho$.

Linear measurements are ubiquitous in signal processing. Let $\vx$ denote some signal of interest, such as an angiogram to be captured by an MRI scanner. One collects linear measurements of the form $\vy = \calA(\vx) + \vn$, where $y_i = \langle \vx, \va_i \rangle + n_i$, $\va_i$ is a measurement pattern (e.g., controlled by the magnetic field of the MRI machine), and $\vn$ is a vector of additive (e.g., Gaussian) noise. The linear measurement model $\vy = \calA(\vx) + \vn$ is very common, not only when $\vx$ represents a physical signal but also in machine learning contexts where $\vx$ represents a collection of data organized into a matrix or tensor. The measurement patterns $\va_i$ vary depending on the context. In some cases, one might collect noisy samples of individual entries of $\vx$; this corresponds to scenarios where each $\va_i$ contains all zeros except for a single entry equal to one. Note, however, that additive Gaussian noise is different from the multinomial noise in QST, although at high sample rates, multinomial noise becomes approximately Gaussian.

Recovering $\vx$ from noisy linear measurements $\vy = \calA(\vx) + \vn$ is known as a linear inverse problem. Such problems are well-studied and can be solved with classical least-squares (i.e., pseudoinverse) techniques when $\dim(\vy) \ge \dim(\vx)$. However, least-squares techniques may not be appropriate in compressive measurement settings where $\dim(\vy) \ll \dim(\vx)$. Rather, because there exist infinitely many $\vx'$ such that $\calA(\vx')=\calA(\vx)$, one must regularize the inverse problem by exploiting an a priori assumption on the structure of $\vx$. In compressive sensing problems~\cite{baraniuk2007compressive}, one often assumes that $\vx$ is sparse, i.e., that it contains relatively few nonzero entries in some orthonormal basis (such as wavelets). In matrix recovery problems~\cite{park2017non,chi2019nonconvex}, one often assumes that $\vx$ has low rank compared to its ambient dimension. In more general machine learning problems, one might assume that $\vx$ is generated according to some neural network or parametric model~\cite{eftekhari2015new,bora2017compressed}. These assumptions restrict $\vx$ to a low-dimensional subset $\setX$ of the ambient signal space. Ideally, then, there will exist no other $\vx' \in \setX$ such that $\calA(\vx')=\calA(\vx)$. Even more strongly, the operator $\calA$ may act as an approximate isometry over the subset $\setX$, meaning it preserves the distances between any two signals in that set. Known as the restricted isometry property (RIP) in compressive sensing, this geometrically corresponds to~$\setX$ having a stable embedding in the measurement space. Such a stable embedding not only guarantees that the information in $\vx$ is preserved in the compressive measurements $\vy$, but it can also be used to prove that structure-aware algorithms can recover accurate approximations of $\vx$ from $\vy$.

Establishing the RIP for a measurement operator $\calA$ and signal family $\setX$ is often facilitated by incorporating randomness into the design of $\calA$. The most classical example in compressive sensing is when each $\va_i$ is populated with independent and identically distributed (i.i.d.) Gaussian entries. Using geometric covering numbers and concentration arguments from the theory of empirical processes~\cite{baraniuk2007simple}, one can then prove that the RIP holds with high probability when the number of measurements scales in proportion to the sparsity level of the signal (times a logarithmic factor). In short, and ignoring the logarithmic factor, one requires only that $\dim(\vy)$ scale with the intrinsic dimension (or degrees of freedom) of $\setX$, which can be far lower than the ambient dimension of $\vx$.

\subsection{Quantum measurement settings}

The specific configurations or arrangements of POVMs used to perform measurements on a quantum system are referred to as measurement settings. These settings are crucial in determining what aspects of the quantum state are being probed and what information can be inferred from the measurement outcomes. Some common choices of measurement settings are included in the following subsections.

\subsubsection{Global measurements}
\label{sec:global-measurement}

\paragraph{Informationally complete POVM: Spherical $t$-designs}

An informationally complete POVM (IC-POVM) is characterized by the property that every quantum state can be uniquely determined from its measurement statistics.
Specifically, the associated linear map $\calA$ is informationally complete if and only if it is injective on the affine space of density operators.
Equivalently, due to the unit trace constraint, this requires $\calA$ to span the $(d^{2n}-1)$-dimensional space of traceless Hermitian operators \cite{acharya2021informationally}.
A notable class of IC-POVMs includes those induced by spherical $t$-designs, particularly when $Q = 1$. Spherical $t$-designs serve as finite sets of points that approximate the entire unit sphere and play a fundamental role in constructing IC-POVMs. To formalize this concept, we provide the definition of spherical $t$-designs below:
\begin{defi}
\label{definition_of_T_Design} (Spherical $t$-designs \cite{matthews2009distinguishability,kueng2017low}). A finite set $\{\vw_k  \}_{k=1}^K\subset \C^{d^n}$ of normalized vectors is called a  spherical quantum $t$-design if \footnote{ $K\geq C_{d^n + \lfloor t/2 \rfloor - 1}^{\lfloor t/2 \rfloor} \cdot C_{d^n + \lceil t/2 \rceil - 1}^{\lceil t/2 \rceil} $ is  necessary to form a spherical $t$-design\cite{scott2006tight,gross2015partial}.}
\begin{eqnarray} \label{the definition of t designs}
     \frac{1}{K}\sum_{k=1}^K (\vw_k\vw_k^\dagger)^{\otimes s}  = \int (\vw\vw^\dagger)^{\otimes s} d\vw
\end{eqnarray}
holds for any $s\leq t$, where the integral on the right hand side is taken with respect to the Haar measure on the
complex unit sphere in $\C^{d^n}$.
\end{defi}
When $s=1$, we have $\frac{1}{K}\sum_{k=1}^K \vw_k\vw_k^\dagger = \int \vw\vw^\dagger d\vw = \frac{1}{d^n}\mId$, and hence $\mA_k = \frac{d^n}{K} \vw_k\vw_k^\dagger, k = 1,\dots, K$ form a rank-one POVM. For simplicity, we call such an induced POVM
\begin{eqnarray} \label{IC-POVM t-designs}
     \mA_{k} = \frac{d^n}{K} \vw_k\vw_k^\dagger, k = 1,\dots, K
\end{eqnarray}
as a $t$-design POVM. In \eqref{the definition of t designs}, we use the widely adopted setting of uniform weights ($1/K$ for each $\vw_k$) to simplify the analysis. Nonetheless, a more general scenario with varied weights can also be explored \cite{dall2014accessible}. In addition, we note that $t$-designs always exist and can, in principle, be constructed for any dimension and any $t$ \cite{seymour1984averaging, bajnok1992construction}. However, in certain cases, these constructions may be inefficient, as they require vector sets of exponential size \cite{hayashi2005reexamination}. As notable examples, \cite{klappenecker2005mutually} demonstrated that mutually unbiased bases and symmetric informationally complete POVMs constitute spherical 2-design POVMs. Additionally, \cite{kueng2015qubit} showed that the set of all stabilizer states in dimension $2^n$ forms a spherical 3-design POVM.

\paragraph{Projective rank-one measurements (non-informationally complete POVM)}

A particular type of POVM that mimics the layout of modern quantum information processing devices is a rank-one POVM of the form $\{ \mA_k \}_{k=1}^{d^n} = \{\vphi_k\vphi_k^\dagger \}_{k=1}^{d^n}$ where $\{\vphi_k \}_{k=1}^{d^n}$ are unit length and orthogonal to each other (i.e., orthonormal). Therefore, each $\mA_k$ is a projection operator onto the corresponding basis vector $\vphi_k $. Upon measuring the state $\vrho$, the probability of getting result $k$ can be rewritten as $
p_k = \trace(\mA_k\vrho) =  \vphi_k^\dagger \vrho \vphi_k$. If we want to perform a projective measurement defined by the orthonormal basis $\{\vphi_k\}$, we can introduce a unitary matrix $\mU = \begin{bmatrix} \vphi_1 & \cdots & \vphi_{d^n} \end{bmatrix}\in\C^{d^n\times d^n}$ and apply $\mU$ on the state $\vrho$ before performing the projective measurement with a physically convenient basis (denoted by $\{\ve_k\}$ below) where $\mU \ve_k = \vphi_k$ for any $k$. Mathematically, this process is written as $p_k = \ve_k^\dagger (\mU^{\dagger} \vrho \mU) \ve_k$.

\subparagraph{Haar random projective measurements}
One might be particularly interested in performing a projective measurement in a random basis, where the unitary matrix $\mU$ is randomly drawn from some ensemble~$\calE$. Different ensembles $\calE$ of accessible unitary transformations give rise to different basis measurement primitives. For instance, when $\calE$ is the unitary group $U(d^n)$ that contains all the possible $d^n\times d^n$ unitary matrices, it is equivalent to randomly sample according to the Haar measure, inducing a global unitary that affects all qudits prior to measuring in the computational basis. A universal quantum computer can approximately generate such random unitary to any given precision, although the number of single and two-qubit quantum gates required in general scale exponentially with the number of qubits  \cite{knill1995approximation}. We call such a projective measurement a {\em Haar random  projective measurement}. Such measurements have been commonly used in optimal quantum state tomography protocols \cite{haah2017sample}, as they typically provide the most unbiased information of an unknown quantum state.

Since a projective measurement in a specific basis only gives $d^n$ linear measurements, it is insufficient to recover a general quantum state. In practice, we can generate multiple such random unitary matrices to measure the state.

\subparagraph{Unitary $t$-designs}
Considering that truly random unitaries drawn from the Haar measure are difficult to implement in practice due to the continuous and infinite nature of the unitary group, unitary $t$-designs provide a structured and experimentally feasible alternative.
Rather than sampling from the entire unitary group, a unitary $t$-design consists of a finite ensemble of unitary matrices whose statistical moments up to order $t$ match those of the Haar distribution.
\begin{defi}
\label{definition_of unitary_T_Design} (Unitary $t$-designs \cite{gross2007evenly,dankert2009exact,roy2009unitary}). 
A set of $H$ unitary matrices $\{ \mU_i \}_{i=1}^{H}\subset \C^{d^n\times d^n}$ is said to be a unitary $t$-design if it satisfies
\begin{eqnarray} \label{the definition of unitary t designs}
     \frac{1}{H}\sum_{i=1}^H \mU_i^{\otimes t}\otimes (\mU_i^\dagger)^{\otimes t}  = \int \mU^{\otimes t}\otimes (\mU^\dagger)^{\otimes t}  d\mU,
\end{eqnarray}
where the integral on the right hand side is taken with respect to the  Haar measure on the unitary group in $\C^{d^n\times d^n}$.
\end{defi}
Exact unitary $t$-designs can be efficiently constructed for small $t$. For example,
 the $n$-qubit Clifford group\footnote{For systems consisting of $n$ qubits, the Clifford group is generated by CNOT, Hadamard, and phase gates. This results in a finite group of cardinality $2^{O(n^2)}$, which maps (tensor products of) Pauli matrices to Pauli matrices under conjugation.} has been shown to serve as an exact unitary 3-design \cite{webb2016clifford,zhu2017multiqubit}, satisfying the statistical properties required for many quantum information tasks \cite{huang2020predicting}. However, higher-order exact designs are generally elusive. To address this, random quantum circuits can be employed to construct approximate unitary $t$-designs in polynomial depth \cite{haferkamp2022random,harrow2023approximate}.

Once a collection of unitary matrices $\{\mU_i\}_{i=1}^{Q}$ is selected from a unitary $t$-design, we can form a set of POVMs comprising $Q$ elements as follows:
\begin{eqnarray} \label{IC-POVM unitary t-designs}
     \mA_{q,k} = \mU_q\ve_k \ve_k^\dagger\mU_q^\dagger, \quad q = 1,\dots, Q, \  k = 1,\dots, d^n,
\end{eqnarray}
where $\{\ve_k\}$ is a physically convenient basis. 

\subsubsection{Local measurements}

In the previous section, we primarily focused on global measurements, which involve simultaneous rotations of the entire system of qudits. While theoretically significant, implementing these measurements using practical quantum circuits presents considerable challenges. In contrast, local measurements \cite{lancien2013distinguishing,brandao2020fast,jameson2024optimal} are more feasible and can be performed efficiently on current quantum hardware. For ease of comparison between different POVMs, we focus on qubits rather than qudits. Specifically, we begin by generating one set of POVMs for each qubit, denoted as $\{\mA_{k_i}^{(i)}\}_{k_i}, i = 1, \dots, n$, for each qubit. Subsequently, we construct a set of POVMs as follows:
\begin{eqnarray}
\label{elements in a set of local POVM}
\{\mA_{k_1}^{(1)} \otimes \mA_{k_2}^{(2)}\otimes \cdots \otimes \mA_{k_n}^{(n)}  \}_{k_1,k_2,\dots,k_n}.
\end{eqnarray}

\paragraph{Local informationally complete POVM} For local IC-POVMs, two commonly used types of local measurements are considered for each qubit: $(i)$ Local symmetric informationally complete (SIC) measurements:
\begin{eqnarray} \label{set of local SIC-POVM}
    \bigg\{\begin{bmatrix}\frac{1}{2}& 0 \\ 0 & 0 \end{bmatrix}, \begin{bmatrix}\frac{1}{6}& \frac{\sqrt{2}}{6} \\ \frac{\sqrt{2}}{6} & \frac{1}{6} \end{bmatrix}, \begin{bmatrix}\frac{1}{6}& \frac{\sqrt{2}}{6}e^{-i\frac{2\pi}{3}} \\ \frac{\sqrt{2}}{6}e^{i\frac{2\pi}{3}} & \frac{1}{6} \end{bmatrix}, \begin{bmatrix}\frac{1}{6}& \frac{\sqrt{2}}{6}e^{-i\frac{4\pi}{3}} \\ \frac{\sqrt{2}}{6}e^{i\frac{4\pi}{3}} & \frac{1}{6} \end{bmatrix}  \bigg\}.
\end{eqnarray}
$(ii)$ Local spherical $3$-design POVM:
\begin{eqnarray} \label{set of local 3 designs}
    \bigg\{\begin{bmatrix}0& 0 \\ 0 & \frac{1}{3} \end{bmatrix}, \begin{bmatrix}\frac{1}{3}& 0 \\ 0 & 0 \end{bmatrix}, \begin{bmatrix}\frac{1}{6}& \frac{1}{6} \\ \frac{1}{6} & \frac{1}{6} \end{bmatrix}, \begin{bmatrix}\frac{1}{6}& -\frac{1}{6} \\ -\frac{1}{6} & \frac{1}{6} \end{bmatrix}, \begin{bmatrix}\frac{1}{6}& \frac{i}{6} \\ -\frac{i}{6} & \frac{1}{6} \end{bmatrix}, \begin{bmatrix}\frac{1}{6}& -\frac{i}{6} \\ \frac{i}{6} & \frac{1}{6} \end{bmatrix}  \bigg\}.
\end{eqnarray}
It should be noted that this set of POVMs is informationally complete only for a single qubit.
For an $n$-qubit system, the tensor product of local POVMs fails to cover the full $(d^{2n}-1)$-dimensional Hilbert-Schmidt space of density operators, and therefore does not constitute a globally informationally complete measurement.

\paragraph{Local Haar random projective measurements} Similarly to non-informationally complete global measurements, a single local POVM is insufficient for recovering the quantum state, even when the measurement is repeated infinitely many times. Therefore, it is necessary to prepare $Q$ distinct POVMs. Specifically, we first construct $n$ unitary matrices $\mU_i \in \C^{2 \times 2}, i = 1, \dots, n$, to form one POVM set. This process is then repeated $Q$ times to generate $Q$ distinct POVMs. In addition, we observe that when $Q \ll 2^n$, the computational complexity of optimization methods using local Haar measurements is significantly lower than that of local IC-POVMs. This difference arises because the total number of measurements for local Haar measurements is $Q2^n$, whereas for local SIC-POVMs and spherical $3$-designs, the numbers are $4^n$ and $6^n$, respectively.

\paragraph{Pauli basis measurements} The efficient implementation of local Haar measurements requires generating random unitaries for each qubit. However, the randomized operations involved can lead to the accumulation of gate errors, which ultimately reduce the fidelity of the measurement outcomes. An alternative approach is the use of Pauli basis measurements \cite{guctua2020fast}, which rely on structured operations and exhibit lower cumulative errors, making them particularly well-suited for noisy intermediate-scale quantum (NISQ) devices. Specifically, each two-dimensional Pauli basis matrix can be expressed as
$\sigma_1  = \mE_1^+ - \mE_1^- = \begin{bmatrix} 0.5 & 0.5 \\ 0.5 & 0.5  \end{bmatrix} - \begin{bmatrix} 0.5 & -0.5 \\ -0.5 & 0.5  \end{bmatrix}$, $\sigma_2  = \mE_2^+ - \mE_2^- = \begin{bmatrix} 0.5 & 0.5i \\ -0.5i & 0.5 \end{bmatrix} - \begin{bmatrix} 0.5 & -0.5i \\ 0.5i & 0.5  \end{bmatrix}$ and $\sigma_3  = \mE_3^+ - \mE_3^- = \begin{bmatrix} 1 & 0 \\ 0  & 0 \end{bmatrix} - \begin{bmatrix} 0 & 0 \\ 0  & 1 \end{bmatrix}$ with three two-outcome POVMs $\{\mE_i^+, \mE_i^-\}, i =1,2,3$ corresponding to its eigenvectors. Thus, there are three potential POVMs for each qubit, resulting in a total of $3^n$ possible local measurements as described in \eqref{elements in a set of local POVM}.

\subsubsection{Pauli observable measurements}

Besides Pauli basis measurements, Pauli observables/matrices are also widely used in quantum measurements. For a single qubit, they are represented by the following $2 \times 2$ matrices:
\begin{eqnarray}
\label{Pauli basis in 2x2}
\sigma_0 = \begin{bmatrix}1 & 0\\  0 & 1 \end{bmatrix},\ \sigma_1 = \sigma_x= \begin{bmatrix}0 & 1 \\ 1 & 0 \end{bmatrix}, \sigma_2 = \sigma_y=\begin{bmatrix}0 & i \\ -i & 0 \end{bmatrix},\ \sigma_3 = \sigma_z=\begin{bmatrix}1 & 0 \\ 0 & -1 \end{bmatrix}.
\end{eqnarray}
For $n$ qubits,  we can obtain $2^n$-dimensional Pauli matrices $\mW_q\in\C^{2^n\times 2^n}$ by forming $n$-fold tensor products of $\sigma_0,\sigma_1,\sigma_2,\sigma_3$:
\begin{eqnarray}
\mW_q =  \sigma_{q_1}\otimes \sigma_{q_2}\otimes \cdots \otimes \sigma_{q_n},  (q_1,\ldots,q_n)\in \{0,1,2,3\}^n,  \forall q = 1, \ldots,4^n.
\label{eq:Pauli-matrix}
\end{eqnarray}
The set of normalized $2^n$-dimensional Pauli matrices, $\frac{1}{\sqrt{2^n}}\{\mW_1, \ldots, \mW_{4^n}\}$, forms an orthonormal basis for $\C^{2^n \times 2^n}$. As a result, any density matrix $\vrho \in \C^{2^n \times 2^n}$ can be uniquely represented by its inner products $\<\mW_1, \vrho \>, \dots, \<\mW_{4^n}, \vrho \> $. These inner products, physically interpreted as the expectation values of the $n$-qubit Pauli observables $\mW_q$, are referred to as Pauli observable measurements. Note that the Pauli matrices
themselves do not form a POVM; thus, it is useful to relate the Pauli matrices to one or more POVMs where quantum measurements may be obtained.

Following \cite{flammia2012quantum,kim2023fast,hsu2024quantum}, each $\mW_q$ can be expressed as a linear combination of local Pauli basis measurement operators.
To set up the subsequent construction, we first note that each two-dimensional Pauli matrix can be decomposed as $\sigma_i = \mE_i^+ - \mE_i^-$ for $i=1,2,3$, where $\mE_i^\pm$ are the corresponding projectors.
In addition,  since $\sigma_0$ is the identity matrix, it can be associated with any rank-one orthonormal POVM elements $\{\mE_i^{\pm}\}$, specifically satisfying $\sigma_0 = \mE_i^+ + \mE_i^-$ for all $i = 1, 2, 3$. To simplify notation, we set $\mE_0^\pm = \mE_3^\pm$. Consequently, each Pauli basis matrix $\sigma_i$ can be uniformly associated with a two-outcome POVM $\{\mE_i^{\pm}\}$ as follows:
\begin{eqnarray}
\label{unified way of Pauli matrix sigma}
\sigma_i = \mE_i^+ + \wh\alpha_i \mE_i^-, \ \forall i = 0,1,2,3, \text{where} \ \wh\alpha_0 = 1, \wh\alpha_1 = \wh\alpha_2 = \wh\alpha_3 = -1.
\end{eqnarray}
Furthermore, we can rewrite the $q$-th $2^n$-dimensional Pauli matrix $\mW_q$ as
\begin{eqnarray}
\label{eq:Pauli-local-POVM}
\mW_q &\!\!\!\!\!\! =\!\!\!\!\!\!&  \sigma_{q_1}\otimes  \cdots \otimes \sigma_{q_n}\nonumber\\
&\!\!\!\!\!\! =\!\!\!\!\!\!&( \mE_{q_1}^+  + \wh\alpha_{q_1} \mE_{q_1}^-) \otimes  \cdots \otimes (\mE_{q_n}^+  + \wh \alpha_{q_n} \mE_{q_n}^-) \nonumber\\
&\!\!\!\!\!\! =\!\!\!\!\!\!& \mE_{q_1}^+ \otimes \mE_{q_2}^+ \otimes \cdots \mE_{q_n}^+ + \wh\alpha_{q_1} \mE_{q_1}^- \otimes \mE_{q_2}^+ \otimes \cdots \otimes  \mE_{q_n}^+ + \cdots + \wh\alpha_{q_1} \mE_{q_1}^-\otimes \wh\alpha_{q_2} \mE_{q_2}^- \otimes \cdots \otimes \wh \alpha_{q_n} \mE_{q_n}^- \nonumber\\
&\!\!\!\!\!\!=:\!\!\!\!\!\!& \sum_{j = 1}^{2^n} \alpha_{q,j} \mE_{q,j}, \quad \alpha_{q,1},\ldots, \alpha_{q,2^n} \in \pm 1,
\end{eqnarray}
which implies that we can define a Pauli basis-based $2^n$-outcome POVM:
\begin{eqnarray}
\label{defi of POVM in Pauli basis measurements}
\bigg\{\mA_{q,1},\ldots,\mA_{q,2^n}\bigg\} = \bigg\{\mE_{q_1}^{\pm} \otimes \cdots \otimes \mE_{q_n}^{\pm}\bigg\}.
\end{eqnarray}
Quantum measurements with this POVM will be characterized by $2^n$ probabilities (population measurements):
\begin{eqnarray}
\label{defi of Pauli basis measurements}
\bigg\{ p_{q,1},\ldots,  p_{q,2^n}\bigg\} = \bigg\{\<\mA_{q,1}, \vrho\>, \ldots, \<\mA_{q,2^n}, \vrho\> \bigg\}.
\end{eqnarray}
Finally, the Pauli observable measurement with $\mW_q$ can be expressed using a linear combination of these probabilities\footnote{{Note that besides the construction above, which is suitable for quantum circuits, any Pauli operator can be written as
\[
\mW_q = \frac{1}{2}(\mId+\mW_q) - \frac{1}{2}(\mId-\mW_q),
\]
where the two terms can each be interpreted as POVM elements. Based on this decomposition, $\mW_q$ naturally defines a two-outcome POVM.
}}:
\begin{eqnarray}
\label{association Pauli measurement and Pauli basis measurements}
\<\mW_q, \vrho \> = \sum_{j=1}^{2^n} \alpha_{q,j} \cdot  p_{q,j}.
\end{eqnarray}

From the discussion above, it follows that we can obtain empirical estimates of $\{\<\mW_q, \vrho \>\}$ by performing the quantum measurements in \eqref{defi of Pauli basis measurements}. Specifically, for a fixed $q \in \{1, 2, \dots, 4^n\}$, the experiment is repeated $M$ times, yielding empirical probabilities $\{\wh p_{q,1},\ldots, \wh p_{q,2^n}\}$. The $q$-th Pauli observable measurement is then estimated as follows:
\begin{eqnarray}
\label{empirical Pauli basis measurements}
 \<\mW_q, \vrho \> \approx \sum_{j=1}^{2^n} \alpha_{q,j} \cdot \wh p_{q,j},
\end{eqnarray}
where the right-hand side is also called an empirical Pauli observable measurement.

\section{Quantum State Tomography}
\label{sec: quantum state tomography}

\subsection{Problem formulation}

Quantum state tomography (QST) aims to reconstruct or estimate the unknown ground-truth density matrix $\vrho^\star \in \C^{d^n \times d^n}$ of a quantum system by performing measurements on repeated copies of the same state. Specifically, given an ensemble of $Q$ POVMs, each repeated for $M$ measurement shots, one obtains empirical measurement statistics $\wh{\vp}$ as in \eqref{eq:map-M-POVM1}. The goal of QST is to construct an estimator $\wh{\vrho}$ from the empirical measurements such that $\wh{\vrho}$ is close to the true state $\vrho^\star$.

The reconstruction error can be quantified using different metrics depending on the application. Common choices include the trace distance $\|\wh{\vrho} - \vrho^\star\|_1$, the Frobenius distance $\|\wh{\vrho} - \vrho^\star\|_F$, and the fidelity $
F(\vrho^\star,\wh{\vrho})
=
\left(
\operatorname{Tr}
\sqrt{
\sqrt{\vrho^\star}
\wh{\vrho}
\sqrt{\vrho^\star}
}
\right)^2,$
which capture different notions of statistical or operational closeness between the estimated and true quantum states.

However, for general quantum states, the required sample complexity $QM$ scales exponentially with the system size $n$, rendering full QST intractable for large-scale systems, regardless of the measurement design \cite{haah2017sample}. This challenge motivates restricting attention to structured quantum states that admit compact representations, such as low-rank states, tensor-network states, neural quantum states, and shallow quantum circuit states. By reducing the effective degrees of freedom, these representations can substantially improve the statistical and computational efficiency of tomography. At the same time, scalable QST depends not only on state structure, but also on measurement design and reconstruction algorithms. Appropriate measurement settings must preserve sufficient information about the structured state while remaining experimentally feasible, whereas efficient algorithms must exploit this structure to recover the state from noisy and incomplete measurement data. These three components---\emph{structure}, \emph{measurement design}, and \emph{algorithm design}---form the central themes of modern structured QST. In the following sections, we first review commonly studied structured state classes, and then survey how these structures influence sample complexity, practical measurement settings, and algorithmic design for scalable tomography.

\subsection{Structured quantum states}
\label{sec:structured-states}

To achieve practical QST for current quantum devices at the scale of $\sim 100$ qubits, it is essential to substantially reduce the sample complexity compared to that required for learning a generic quantum state. This reduction is possible only when the target state possesses additional structure, allowing it to admit a compact representation with far fewer effective degrees of freedom than a fully generic $n$-qubit state. Fortunately, many physical quantum states indeed have such structure. Examples include ($i$) states that are known to be classically simulable, such as stabilizer states \cite{englbrecht2020symmetries}, and phase states where the  classical low-degree Boolean functions are encoded as the amplitudes of quantum states \cite{ji2018pseudorandom}, ($ii$) ground states of most quantum systems with short-range interactions and states generated by such quantum systems in a finite amount of time \cite{eisert2010}, such as the matrix product state/operator (MPS/MPO),  ($iii$) neural quantum states that represent complex quantum states using techniques inspired by neural networks, and $(iv)$ states generated by shallow quantum circuits.  Below we provide an introduction to each of them.

\subsubsection{Low-rank states}
For pure or nearly pure quantum states, one inherent structure is characterized by low entropy resulting in a low-rank density matrix~\cite{kueng2017low,guctua2020fast,francca2021fast,voroninski2013quantum,haah2017sample}. When we assume that the density matrix $\vrho$ has rank $r$, we may use a Burer-Monteiro type decomposition \cite{burer2003nonlinear, burer2005local} to represent the density matrix as follows:
    \begin{eqnarray}
    \label{The low-rank density matrix}
    \vrho = \mU {\mU^\star}, \ \text{where} \ \mU\in\C^{d^n\times r} \ \text{and} \ \| \mU\|_F = 1.
    \end{eqnarray}
    For a given density matrix, \eqref{The low-rank density matrix} simultaneously ensures physical validity and enforces low-rank structure through an appropriately chosen small $r$. Compared to a general density matrix, the degrees of freedom for a low-rank matrix are reduced to $O(d^n r)$, indicating that the sampling and computational complexities can also be reduced to this level, as demonstrated later. 

\subsubsection{Tensor-network representations}

\paragraph{Matrix product operators:} Although the computational and storage demands for low-rank density matrices are substantially lower than those for general ones, they still scale exponentially with the number of qudits, $n$. Furthermore, the high-purity assumption underlying the low-rank approximation becomes increasingly unrealistic for current processors in the noisy intermediate-scale quantum (NISQ) era. This underscores the need to explore alternative assumptions for reducing the parameter count. For example, ground states of quantum systems with short-range interactions, along with states produced by such systems over finite durations \cite{eisert2010}, frequently exhibit entanglement localized within subsystems of a one-dimensional noisy quantum computer \cite{noh2020efficient}. This localization enables a compact representation using matrix product states (MPSs) for vector forms or matrix product operators (MPOs) for matrix forms.

\begin{figure}[!ht]
\centering
\includegraphics[width=14.5cm, keepaspectratio]%
{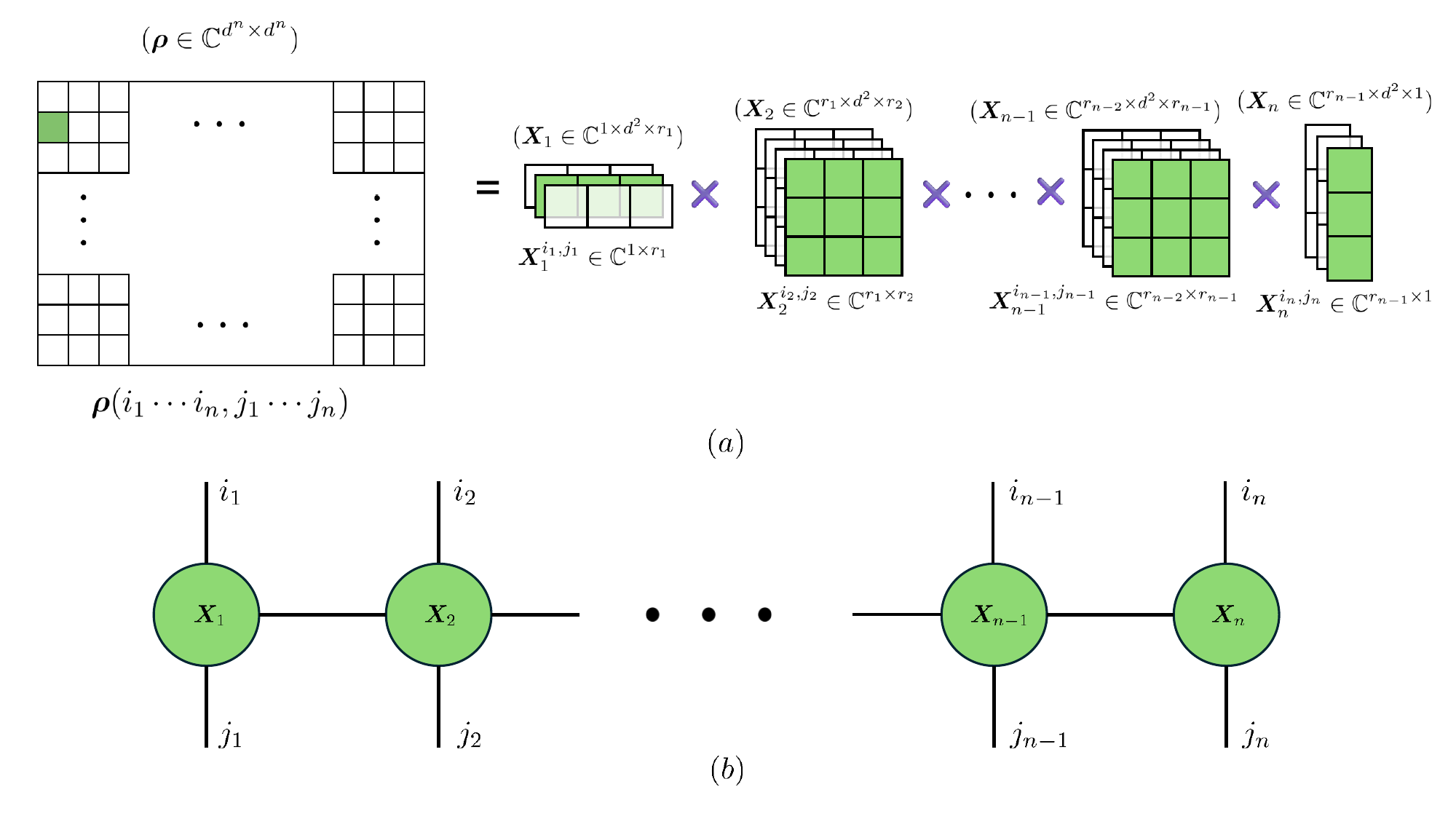}
\vspace{-0cm}
\caption{Illustration of the MPO in \eqref{DefOfMPOe} from two perspectives: (a) each entry of the density matrix can be represented as products of $n$ matrices, where green represents one entry and the corresponding $n$ matrices, and (b) each element of the density matrix is illustrated in a diagrammatic form, where the line connecting two circles signifies the tensor contraction operation \cite{cichocki2014tensor}, and unconnected line segments denote indices.}
\label{TheMPOfig}
\end{figure}

    For a density matrix $\vrho\in\C^{d^\nqbit\times d^ \nqbit}$ representing an $\nqbit$-qudit quantum system, the indices of rows (columns) are specified using a single index-array $i_1\cdots i_\nqbit$ ($j_1\cdots j_\nqbit$), where $i_1,\ldots,i_\nqbit\in[d]$.\footnote{ Specifically, $i_1\cdots i_n$ represents the $(i_1+\sum_{\ell=2}^nd^{\ell-1}(i_\ell-1))$-th row.} The matrix $\vrho$ is referred to as an MPO if its $(i_1\cdots i_\nqbit,j_1\cdots j_\nqbit)$-element can be expressed as a matrix product~\cite{werner2016positive}
    \begin{eqnarray}
    \label{DefOfMPOe}
    \vrho(i_1 \cdots i_\nqbit, j_1 \cdots j_\nqbit)  =  \mX_1^{i_1,j_1} \mX_2^{i_2,j_2} \cdots \mX_\nqbit^{i_\nqbit,j_\nqbit},
    \end{eqnarray}
    where $\mX_\ell^{i_\ell,j_\ell}\in \C^{r_{\ell-1}\times r_\ell}$ with $r_0 = r_\nqbit = 1$. An illustration is provided in Figure~\ref{TheMPOfig}. The dimensions $\vr = (r_1,\ldots,r_{n-1})$, commonly referred to as the {\it bond dimensions} in quantum physics, are also known as the {\it MPO ranks}. These dimensions are associated with the ranks of specific matrices obtained by reshaping the density matrix  $\vrho$ in various ways. An MPO comprises $nd^2$ matrices, each with a dimension of at most $\ol r \times \ol r$ where the matrix dimension
    \begin{eqnarray}
    \label{bond dimension for MPO}
    \ol r = \max_\ell r_\ell
    \end{eqnarray}
    determines the efficiency of the representation. Thus, an MPO can be efficiently represented with $O(nd^2\ol r^2)$ parameters. In addition, it is worth noting that an MPO is equivalent to a tensor train (TT) representation commonly used for describing high-dimensional tensors in machine learning~\cite{qin2024quantum}. Consequently, the tensor train singular value decomposition (TT-SVD) \cite{oseledets2011tensor} can always be employed to derive its canonical form  in which all the bond indices simultaneously correspond to orthonormal Hilbert spaces.

\paragraph{Projected entangled-pair operators:} MPOs are extensively utilized for the study of one-dimensional quantum systems; however, their extension to higher dimensions poses significant challenges, particularly in analyzing interacting spin systems on two-dimensional lattices. A common approach involves adapting MPOs by arranging spins linearly across a two-dimensional lattice. While effective in certain cases \cite{yan2011spin}, this method often fails to capture the intricate entanglement structures typical of two-dimensional ground states, especially as system size increases. To address these limitations, an alternative representation---projected entangled-pair operators (PEPOs)---has been proposed. PEPOs represent pairs of maximally entangled auxiliary states confined within low-dimensional subspaces, offering a more flexible framework for higher-dimensional systems. Importantly, PEPOs have been shown to effectively describe various physical states, including the 2D cluster state \cite{raussendorf2001one}, the Toric Code model \cite{kitaev2003fault}, the 2D resonating valence bond state \cite{anderson1987resonating}, and the 2D AKLT model \cite{affleck1988valence,wei2011affleck}. Comprehensive discussions and illustrative examples of PEPO structures in two dimensions are available in \cite{cirac2021matrix}.
    \begin{figure}[!ht]
    \centering
    \includegraphics[width=14.5cm, keepaspectratio]%
    {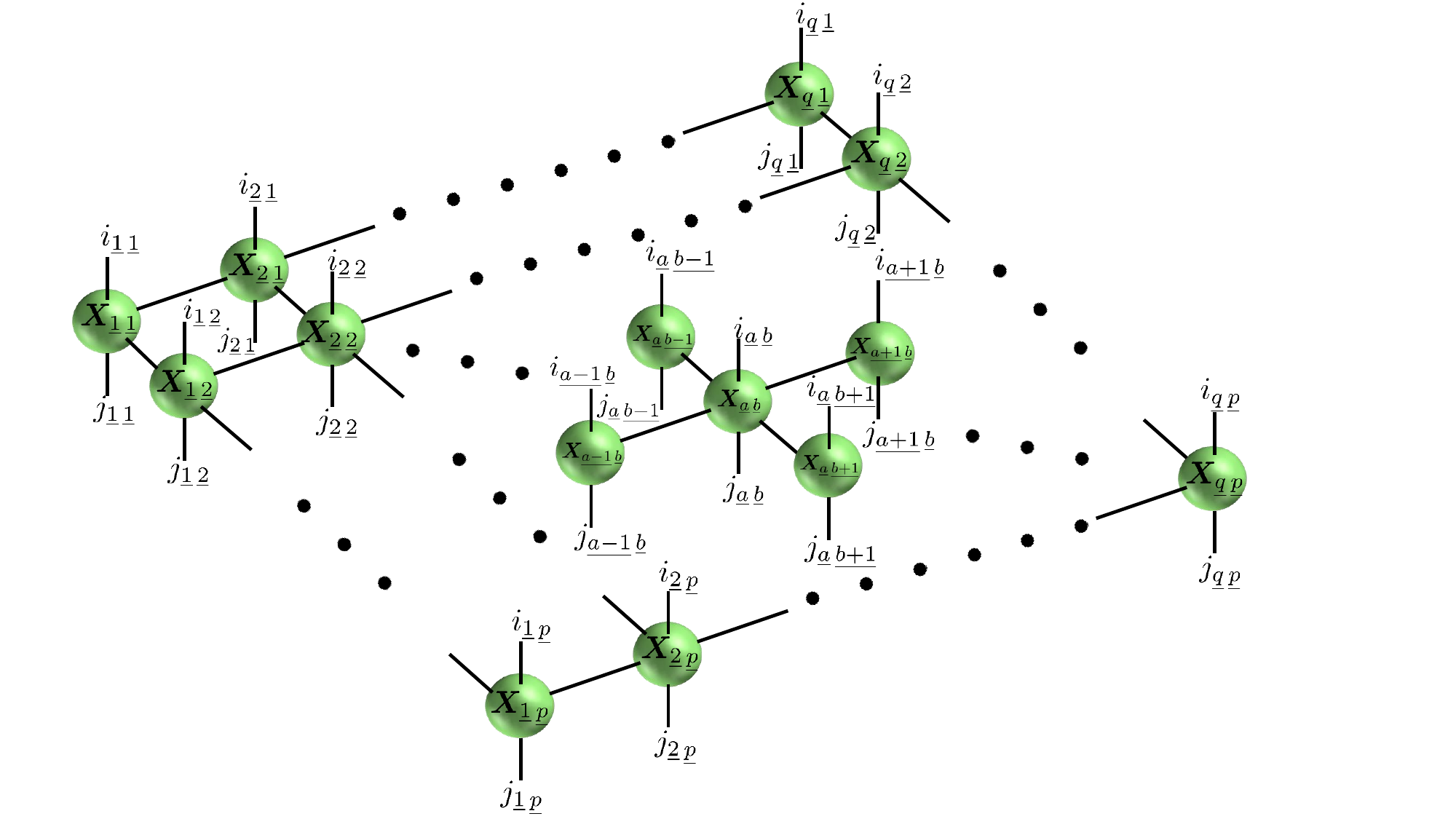}
    \vspace{-0cm}
    \caption{Illustration of the PEPO  from each element of the density matrix is illustrated in a diagrammatic form, where the line connecting two circles signifies the tensor contraction operation \cite{cichocki2014tensor}, and unconnected line segments denote indices.}
    \label{ThePEPOfig}
    \end{figure}

For a density matrix $\vrho \in \C^{d^{\nqbit}\times d^{\nqbit}}$ in an $\nqbit$-qudit quantum system, with $n = qp$ for positive integers $q$ and $p$, the PEPO representation consists of local core tensors
    $\mX_{\ul{a}\,\ul{b}}^{\,i_{\ul{a}\,\ul{b}},\,j_{\ul{a}\,\ul{b}}}$,
    for $a \in [q]$ and $b \in [p]$.
    Each core tensor is a four-way complex array with dimensions
    $r_{\ul{a}\,\ul{b-1},\,\ul{a}\,\ul{b}} \times r_{\ul{a-1}\,\ul{b},\,\ul{a}\,\ul{b}} \times r_{\ul{a}\,\ul{b},\,\ul{a}\,\ul{b+1}} \times r_{\ul{a}\,\ul{b},\,\ul{a+1}\,\ul{b}}$,
    subject to boundary conditions
    $r_{\ul{a}\,\ul{0},\,\ul{a}\,\ul{1}} =r_{\ul{0}\,\ul{b},\,\ul{1}\,\ul{b}} =r_{\ul{a}\,\ul{p},\,\ul{a}\,\ul{p+1}} =
    r_{\ul{q}\,\ul{b},\,\ul{q+1}\,\ul{b}} = 1$.
     See Figure~\ref{ThePEPOfig} for an illustration. Detailed definitions can be found in \cite{qin2025quantum}. Similar to the MPO format, the degrees of freedom for a PEPO are given by $O(nd^2\wt r^4)$ with a bond dimension
     \begin{eqnarray}
     \label{Bond diemsion for PEPO}
         \wt r = \max_{i,j,k,l}r_{\ul{i}\,\ul{j},\ul{k}\,\ul{l}}.
     \end{eqnarray}
     This scaling remains linear with respect to the number of qudits $n$, significantly reducing the storage requirements compared to a general quantum state. In addition, we note that an MPO can be regarded as a special case of a PEPO without loops, effectively representing a single column or row in the PEPO structure. However, as highlighted in \cite{orus2014practical}, unlike MPOs with open boundary conditions, PEPOs lack a canonical form. Specifically, it is generally impossible to simultaneously choose orthonormal bases for all indices within the PEPO network. This limitation arises because the presence of a loop in the tensor network prevents a formal division of the network into distinct left and right pieces by simply cutting one index. Consequently, orthonormal bases (i.e., Schmidt bases) cannot be consistently defined for the left and right subsystems relative to a given index, precluding the definition of a canonical form in this context.

\paragraph{Tensor networks for state vectors:} Different from MPOs and PEPOs which can conveniently represent mixed states/density matrices as pointed out in \cite{orus2019tensor}, many tensor networks are mainly used to compactly represent the state vector/wave function $\vpsi$ in the pure state $\vrho = \vpsi\vpsi^\dagger$ including the matrix product state (MPS), the projected entangled-pair state (PEPS), the tree tensor network (TTN) \cite{shi2006classical} and the multiscale entanglement renormalization ansatz (MERA) \cite{vidal2007entanglement}. MPS and PEPO are special forms of MPO and PEPO when $j_1\cdots j_n = 1$. The TTN adopts a hierarchical tree-like structure, making it particularly effective for representing states with low entanglement, especially those characterized by a hierarchical entanglement pattern. The MERA extends the TTN by incorporating disentanglers, enabling it to efficiently capture entanglement across multiple length scales, making it particularly well-suited for critical quantum states and systems with scale invariance. This review primarily focuses on compact tensor representations for mixed states rather than state vectors. For comprehensive discussions on tensor network techniques, we refer readers to \cite{orus2014practical,orus2014advances,orus2019tensor,acuaviva2023minimal}, while computational methods for tensor networks are summarized in \cite{orus2019tensor}.

\subsubsection{Neural quantum states}
Neural quantum states have been extensively utilized to represent the state vector or wave function $\vpsi$ in the context of a pure state, where the density matrix is given by $\vrho = \vpsi\vpsi^\dagger$:
    \begin{eqnarray}
    \label{DefOfneural quantum state}
    \vpsi(i_1 \cdots i_\nqbit)  =  h_{\vtheta}(i_1,\dots, i_n),\ i_1,\dots, i_n\in[d],
    \end{eqnarray}
    and where $h_{\theta}(i_1,\dots, i_n)$ represents a neural network model with parameters $\vtheta$. Various neural network architectures have been utilized to represent neural quantum states, including restricted Boltzmann machines \cite{torlai2018neural,torlai2019integrating}, feedforward neural networks \cite{cai2018approximating}, recurrent neural networks \cite{morawetz2021u,iouchtchenko2023neural}, transformer networks \cite{cha2021attention,ma2023tomography}, convolutional neural networks, and variational autoencoders \cite{rocchetto2018learning}. A detailed summary of these architectures and their applications can be found in \cite{lange2024architectures}. However, compared to low-rank or tensor network models, which possess well-defined structures for $n$ qudits, neural quantum states face two primary challenges: $(i)$ the choice of nonlinear operations, such as activation functions, is influenced by factors including the characteristics of the input data, network depth, optimization objectives, and computational efficiency, making the practical design process more intricate, and $(ii)$ the degrees of freedom in these networks are not explicitly defined. Although increasing the number of parameters in the network may enhance representational accuracy, when the parameter count exceeds the intrinsic degrees of freedom $d^n$ in the wave function, neural quantum states risk becoming non-compact representations. Given the variety of available network models, determining which quantum states are suitable for representation as corresponding neural quantum states remains an open question.

\subsubsection{Shallow quantum circuits:}
Besides the structured quantum states discussed above, shallow quantum circuits \cite{huang2024learning,kim2024learning} have also been employed to represent quantum states. Specifically, the pure state  $\vrho = \vpsi\vpsi^\dagger$ can be written as
    \begin{eqnarray}
    \label{DefOf shallow quantum circuit}
    \vpsi   =  \mU(\vtheta) \ve_1 = \Pi_{\ell = 1}^{L}\mU_\ell(\vtheta_\ell)\ve_1
    \end{eqnarray}
    where $\ve_1 = \begin{bmatrix}1 & 0 & \cdots & 0  \end{bmatrix}^\top\in\R^{d^n\times 1}$ denotes the initial state, $L$ is the circuit depth, $\mU_\ell(\vtheta_\ell)$ denotes the local parameterized gates at the  $\ell$-layer (typically single- or two-qudit gates), and $\vtheta = \{\vtheta_\ell\}_{\ell=1}^{L}$ are the circuit parameters. The state is therefore specified compactly by the parameter set $\vtheta = \{\vtheta_\ell\}_{\ell=1}^{L}$, without the need to explicitly store all amplitudes. Furthermore, if the circuit is shallow (small $L$) and local, the reduced density matrix of each qudit depends only on $O(L)$ neighboring qudits. This locality underpins the efficiency of learning shallow circuits, as it avoids the exponential sample complexity otherwise required.

\subsubsection{Unified notation and geometric complexity}
\label{sec:unified-notation-structure}
To provide a unified presentation for the structured state classes described above, we let $\setX \subset \C^{d^{\nqbit}\times d^{\nqbit}}$ denote the class of quantum states of interest, such as low-rank states, tensor-network states, neural quantum states, or shallow quantum circuit states. Throughout this review, we adopt this unified notation whenever possible, while specifying the precise form of $\setX$ when discussing results for each particular state class. Note that $\setX$ may also represent the set of general physical states, in which case the only constraints are positive semidefiniteness (PSD) and unit trace.

\paragraph{Geometric complexity.}
To provide a unified complexity analysis, we use tools from $\epsilon$-net and covering number theory to characterize the geometric complexity of state classes in $\setX$. Define the normalized set $
\ol{\setX}
=
\left\{
\frac{\vrho}{\|\vrho\|_F}
:\;
\vrho \in \setX
\right\},$
which scales all elements in $\setX$ to have unit Frobenius norm. For $\varepsilon > 0$, a subset $\ol{\setX}_\varepsilon \subset \ol{\setX}$ is called an $\varepsilon$-net (or $\varepsilon$-cover) of $\ol{\setX}$ if, for every
$\frac{\vrho}{\|\vrho\|_F} \in \ol{\setX}$, there exists
$\frac{\vrho'}{\|\vrho'\|_F} \in \ol{\setX}_\varepsilon$
such that
$
\left\|
\frac{\vrho}{\|\vrho\|_F}
-
\frac{\vrho'}{\|\vrho'\|_F}
\right\|_F
\le \varepsilon.
$
The minimum cardinality of such an $\varepsilon$-net is called the covering number of $\setX$, denoted by $\calN_\varepsilon(\setX)$. Intuitively, the covering number is the minimum number of Frobenius balls of radius $\varepsilon$ needed to cover the normalized set $\ol{\setX}$. Covering arguments are useful for controlling the complexity of large or uncountable state classes: rather than directly analyzing all elements in $\setX$, one can instead analyze the finite set $\ol{\setX}_\varepsilon$ and extend the result to the entire class through approximation. The choice of $\varepsilon$ depends on the state structure. For example, $\varepsilon = 1/2$ can be used for low-rank states, while $\varepsilon = 1/(2n)$ is used for MPOs \cite{qin2024}. Since it is often chosen as a constant, in the sequel, we suppress the dependence on $\varepsilon$ and simply write $\calN(\setX)$. The complexity of several structured state classes, characterized through the logarithm of the covering number, is summarized as
\begin{eqnarray}
\label{covering number of different structures}
\log \calN(\setX)
=
\begin{cases}
O(d^{2n}), & \textup{Physical},\\
O(d^n r), & \textup{Low-rank},\\
O(nd^2 \bar r^2 \log n), & \textup{MPO},\\
O(nd^2 \widetilde r^4 \log n), & \textup{PEPO},
\end{cases}
\end{eqnarray}
where $\bar r$ and $\widetilde r$ denote the bond dimensions of the MPO and PEPO representations defined in \eqref{bond dimension for MPO} and \eqref{Bond diemsion for PEPO}, respectively.

\paragraph{Parameterized representation.}
All structured state classes described above admit parameterized representations. For example, low-rank states can be represented through the factor $\mU$ in \eqref{The low-rank density matrix}; MPOs through the local tensor factors $\{\mX_{\ell}^{i_\ell,j_\ell}\}$ in \eqref{DefOfMPOe}; neural quantum states through network parameters $\vtheta$ in \eqref{DefOfneural quantum state}; and shallow quantum circuits through circuit parameters $\vtheta$ in \eqref{DefOf shallow quantum circuit}. To unify these cases, we write $\vrho_{\vtheta}$ to denote a parameterized representation of a quantum state $\vrho$, where $\vtheta$ denotes the corresponding model parameters. We then consider explicit parameterizations of the form
\e
\setX
=
\{
\vrho_{\vtheta}
:
\vtheta \in \Theta
\}.
\label{eq:parameterized-representation}\ee
In this setting, the parameter dimension $\dim(\Theta)$ (which will also be referred to as $\dim(\setX)$ for the ease of presentation) can be viewed as the intrinsic dimension of the model class and provides an alternative measure of complexity for structured quantum states, complementary to geometric measures such as covering numbers.

\subsection{The role of structure in sample complexity and measurement design}

Before reviewing detailed results on exploiting structure for efficient quantum learning, we highlight two complementary goals:

\begin{itemize}
\item \textbf{Fundamental limits:} Reduce the intrinsic sample complexity of the estimation problem by restricting attention to structured state classes.
\item \textbf{Measurement design:} Construct measurement schemes that are experimentally feasible with current quantum devices and achieve near-optimal sample complexity, ideally matching the fundamental limits.
\end{itemize}

The first goal is information-theoretic and characterizes what is achievable in principle. The second is partly algorithmic and focuses on how to attain this performance using practical measurement designs. In this survey, we will see that these two goals are tightly connected through the notion of stable embeddings.

\subsubsection{Role of structure in reducing sample complexity}

A key reason structured quantum states enable efficient tomography is that they admit a \emph{low-dimensional parameterization} in \eqref{eq:parameterized-representation}. Specifically, consider a local parameterization $\vrho_{\vtheta}$ of $\setX$ around the ground truth $\vrho^\star$, where $\vtheta \in \R^{\dim(\setX)}$ and $\dim(\setX) \ll d^{2n}$ is the intrinsic dimension of $\setX$. Estimating $\vrho^\star$ is equivalent to estimating~$\vtheta^\star$. Given $M$ independent measurements for a single POVM, any unbiased estimator $\wh{\vtheta}$ has covariance $\mSigma = \operatorname{Cov}(\wh{\vtheta})$ bounded by the \emph{Quantum Cram\'{e}r--Rao bound} (QCRB) \cite{braunstein1994statistical}:
\[
\mSigma \succeq \frac{1}{M} \calF^{-1}(\vtheta^\star),
\]
where $\calF(\vtheta^\star)$ is the quantum Fisher information matrix (QFIM), which defines the local (Bures) geometry of the state manifold.

To connect this bound to estimation error, consider the infidelity $1 - F(\vrho^\star, \wh{\vrho})$. For small errors, it admits the local approximation
\[
1 - F(\vrho^\star, \wh{\vrho})
\;\approx\;
\frac{1}{4} (\wh{\vtheta} - \vtheta^\star)^\top
\calF(\vtheta^\star)
(\wh{\vtheta} - \vtheta^\star).
\]
Taking expectations and applying the QCRB yields
\[
\mathbb{E}[1 - F(\vrho^\star, \wh{\vrho})]
\;\gtrsim\;
\frac{1}{4M}\trace\!\big(\calF(\vtheta^\star)\calF^{-1}(\vtheta^\star)\big)
=
\frac{\dim(\setX)}{4M}.
\]

This bound shows that the estimation error is governed by the \emph{intrinsic dimension} $\dim(\setX)$ rather than the ambient dimension $d^{2n}$.
This establishes a fundamental (necessary) limit: structure reduces the \emph{statistical difficulty} of quantum state estimation. However, achieving this limit requires appropriate measurement design, which we discuss next.

\subsubsection{Design of efficient POVMs via information-geometric stable embeddings}

We now turn to the second goal: designing measurement schemes that are both experimentally feasible and statistically efficient. A central question in QST is whether the chosen measurement scheme $\calA$ preserves enough information to distinguish different states in the structured class $\setX$. Since the experimental data are generated by repeated sampling from outcome distributions induced by $\vrho$, it is natural to study this question through an information-geometric lens. 

We will use a single POVM to illustrate the main idea, but the analysis can be easily extended to multiple POVMs. For a single POVM $\{\mA_k\}_{k=1}^K$, the state $\vrho$ induces the probability vector
\[
\calA(\vrho)
= \begin{bmatrix}
\<\mA_1,\vrho\> & \cdots & \<\mA_K,\vrho\>
\end{bmatrix}^\top \in \Delta^{K-1},
\]
where $\Delta^{K-1}$ denotes the probability simplex. Repeating this POVM measurement $M$ times yields counts
\[
\vf = (f_1,\ldots,f_K) \sim P_{\vrho,\calA}:=\operatorname{Multinomial}(M,\calA(\vrho)).
\]

For two candidate states $\vrho_1,\vrho_2\in\setX$, a natural way to quantify their distinguishability under the measurement design $\calA$ is via the Kullback--Leibler divergence
\[
D_{\mathrm{KL}}\!\left(P_{\vrho_1,\calA}\,\|\,P_{\vrho_2,\calA}\right)
=
D_{\mathrm{KL}}\!\left(
\operatorname{Multinomial}(M,\calA(\vrho_1))
\,\middle\|\,
\operatorname{Multinomial}(M,\calA(\vrho_2))
\right).
\]
The measurement design $\calA$ is informative for the state class $\setX$ if distinct states induce well-separated outcome distributions.

This motivates the following notion: the map $\calA$ provides an \emph{information-geometrically stable embedding} of $\setX$ if there exist constants $0<c\le C<\infty$ such that for all $\vrho_1,\vrho_2\in\setX$,
\e
c\, d(\vrho_1,\vrho_2)^2
\;\le\;
\frac{1}{M}D_{\mathrm{KL}}\!\left(P_{\vrho_1,\calA}\,\|\,P_{\vrho_2,\calA}\right)
\;\le\;
C\, d(\vrho_1,\vrho_2)^2,
\label{eq:stable-embedding-KL}\ee
where $d(\cdot,\cdot)$ is a suitable metric on quantum states, such as the Frobenius norm, trace norm, or Bures distance. The lower bound ensures identifiability and robustness, while the upper bound controls distortion induced by the measurement map.

When the number of shots per POVM is large, the multinomial model admits a local Gaussian approximation. In that regime, the empirical frequency vector $\wh\vp=\vf/M$ is approximately distributed as
\[
\wh\vp
\approx
\mathcal{N}\!\left(\calA(\vrho),\,\frac{1}{M}\Sigma(\vrho)\right),
\]
where $\Sigma(\vrho)=\diag(\calA(\vrho))-\calA(\vrho)\calA(\vrho)^\top$ is the multinomial covariance matrix. Under this approximation, the KL divergence between two models admits a quadratic expansion. For heuristic simplicity, if we further approximate $\Sigma(\vrho)$ by the identity matrix, the KL divergence reduces to
\e
D_{\mathrm{KL}}\!\left(P_{\vrho_1,\calA}\,\|\,P_{\vrho_2,\calA}\right)
\approx
\frac{M}{2}
\|\calA(\vrho_1)-\calA(\vrho_2)\|_2^2.
\label{eq:KL-L2}\ee
This approximation makes explicit the connection between statistical distinguishability and stable embedding properties of the measurement operator. In particular, it suggests that proving injectivity or restricted isometry properties of $\calA$ over $\setX$ is closely related to establishing favorable estimation guarantees for structured QST.

Plugging the approximation \eqref{eq:KL-L2} into \eqref{eq:stable-embedding-KL}, we obtain a \emph{stable embedding} of the structured state set $\setX$ under the measurement map $\calA$ that is closely related to the classical \emph{restricted isometry property} (RIP) studied in compressive sensing~\cite{donoho2006compressed,candes2006robust,
candes2008introduction,recht2010guaranteed,eftekhari2015new}. In its standard form, the RIP requires that a linear measurement operator approximately preserves the Euclidean norm of all signals in a low-complexity set (e.g., sparse vectors or low-rank matrices). More precisely, when $d(\cdot,\cdot)$ is chosen as the Frobenius norm, the above inequality reduces to a matrix RIP-type condition:
\[
(1-\delta)\|\vrho_1-\vrho_2\|_F^2
\;\le\;
\|\calA(\vrho_1)- \calA(\vrho_2)\|_2^2
\;\le\;
(1+\delta)\|\vrho_1-\vrho_2\|_F^2,
\]
for all $\vrho_1,\vrho_2 \in \setX$, where $\delta \in (0,1)$ is a small constant. This is directly analogous to the RIP for low-rank matrix recovery.

However, there is an important conceptual distinction. In classical compressive sensing, one directly observes (possibly subsampled) linear measurements of the form $\calA(\vrho)$. In contrast, quantum measurements only provide \emph{samples} from the probability distribution induced by $\calA(\vrho)$, introducing an additional layer of statistical noise. The Gaussian approximation above bridges this gap by showing that, in the large-sample regime, the statistical estimation problem behaves similarly to a noisy linear inverse problem with measurement operator $\calA$. Therefore, establishing RIP-like properties for quantum measurement ensembles (e.g., Pauli measurements, random unitary designs, or classical shadows) plays an important role in guaranteeing stable and sample-efficient reconstruction of structured quantum states.

In summary, structure reduces the intrinsic degrees of freedom (via the QCRB), while stable embedding-based measurement design ensures these degrees of freedom can be efficiently recovered in practice.

\subsection{Sample complexity for structured quantum states with IC-POVM measurements}

We first review results for informationally complete POVMs (IC-POVMs), particularly those induced by spherical $t$-designs (\Cref{definition_of_T_Design}), where full state identifiability is guaranteed in the absence of statistical noise from finite sampling. We then turn to more compressive random measurement schemes, which use substantially fewer measurement outcomes, in the next section.

To characterize the sample complexity with IC-POVMs, we first establish an exact isometry property of the associated measurement operator, which serves as a stronger analogue of the standard RIP condition.
\begin{lemma}(\cite[Lemma 2]{qin2024sample})
\label{l2 norm of approximate 2-designs RIP}
Suppose that $\{\vw_k  \}_{k=1}^K\subset \C^{d^n}$ forms a spherical $t$-design ($t\ge 2$). Let $\calA$ be the linear map  corresponding to the induced POVM $\{\mA_k = \frac{d^n}{K} \vw_k\vw_k^\dagger\}$.
Then for arbitrary physical states $\vrho_1, \vrho_2\in\C^{d^n\times d^n}$, $\calA(\vrho_1) - \calA(\vrho_2)$ satisfies
\begin{eqnarray}
\label{The l2 norm of A(rho) approximate 2_designs}
\|\calA(\vrho_1) - \calA(\vrho_2) \|_2^2 = \sum_{k=1}^K\<\mA_k,  \vrho_1 - \vrho_2 \>^2 = \frac{d^n \|\vrho_1 - \vrho_2\|_F^2}{K(d^n + 1)}.
\end{eqnarray}
\end{lemma}
In contrast to the conventional RIP, which provides only upper and lower bounds and only for certain vectors/matrices, \Cref{l2 norm of approximate 2-designs RIP} establishes an exact identity and holds for all physical states. This reflects the fact that spherical $t$-designs ($t\ge 2$) preserve the Hilbert--Schmidt geometry of quantum states under the measurement map. Motivated by this exact isometry property, given empirical measurement outcomes $\wh{\vp}$, we consider the constrained least-squares estimator:
\begin{equation}
    \label{The loss function in QST for IC-POVM general}
    \wh{\vrho} = \argmin_{\vrho\in\setX}\|\calA(\vrho) - {\wh\vp}\|_2^2
    = \argmin_{\vrho\in\setX} \sum_{k=1}^{K}\big(\<\mA_k, \vrho \> - \wh p_k \big)^2.
\end{equation}
The exact isometry property enables a unified recovery analysis for broad classes of structured quantum states. By combining the concentration analysis of empirical measurements \cite{qin2024sample,qin2025quantum} with the geometric complexity of the constraint set $\setX$, the estimation error of \eqref{The loss function in QST for IC-POVM general} admits the following general characterization.
\begin{theorem} (\cite{qin2024sample,qin2025quantum})
\label{Recovery error of various quantum POVM}
Suppose $\{\mA_1,\ldots,\mA_K\}$ is a spherical $t$-design POVM with $t\ge 2$, and let the true state $\vrho^\star\in\setX\subset\C^{d^n\times d^n}$. Also suppose we obtain empirical measurement outcomes $\wh{\vp}$ from $M$ independent measurements of $\vrho^\star$ under this POVM, with
\begin{eqnarray}
M \gtrsim\frac{\log(\calN(\setX))\gamma_t(\vrho^\star)}{\epsilon^2},
\end{eqnarray}
where $\calN(\setX)$ denotes the complexity measure of the state class (i.e., covering number) as described in \Cref{sec:unified-notation-structure}, and $
\gamma_t(\vrho^\star)=
\begin{cases}
\gamma(\vrho^\star), & t=2,\\
1, & t>2,
\end{cases} $ where $\gamma(\vrho^\star)
:=
K\max_k p_k
=
K\max_k\langle \mA_k,\vrho^\star\rangle$.
Then with high probability, the constrained least-squares estimator $\wh{\vrho}$ defined in \eqref{The loss function in QST for IC-POVM general} satisfies
\begin{eqnarray}
\label{final conclusion of recovery error1}
\|\wh{\vrho}-\vrho^\star\|_F
\le \epsilon.
\end{eqnarray}

\end{theorem}
\Cref{Recovery error of various quantum POVM} provides a unified characterization of sufficient conditions for stable QST under the Frobenius norm, explicitly showing that the required sample complexity is governed by the structural complexity of the underlying quantum state model. This perspective closely aligns with the principles of compressive sensing and structured recovery, where reconstruction guarantees are often determined by intrinsic model complexity and covering-number-based arguments. For general physical states, the required number of copies scales exponentially with the number of qudits $n$, reflecting the exponential growth of the ambient Hilbert space dimension. In contrast, for structured state classes such as MPO and PEPO states, the sample complexity scales only polynomially with $n$. This improvement stems from their compact parameterizations: the effective degrees of freedom of MPO states scale linearly with $n$. These results demonstrate the statistical advantage of exploiting low-dimensional quantum state structures in tomography.
Beyond the dependence on the state complexity, these results further illustrate the advantage of IC-POVMs as measurement schemes. The exact isometry property induced by spherical $t$-designs preserves the Hilbert--Schmidt geometry of quantum states, which is significantly stronger than conventional RIP-type guarantees. As a result, IC-POVMs provide a more structured and information-efficient measurement framework, yielding unified and near-optimal recovery guarantees across broad classes of quantum states.

\subsection{Sample complexity for structured quantum states with random measurements}
While the exact isometry property induced by spherical $t$-design POVMs holds uniformly over all quantum states and enables strong recovery guarantees for different structured states, this level of informational completeness often comes at the cost of experimental complexity, making IC-POVMs difficult to implement on current large-scale quantum devices. In many practical settings, one is willing to relax exact isometry for arbitrary states in exchange for measurement schemes that are experimentally simpler and sufficient for structured quantum states. Motivated by this trade-off, randomized measurement frameworks have recently emerged as an important alternative. Although these measurements are generally not informationally complete for the entire state space, they often preserve sufficient geometric information for structured state classes, enabling stable recovery with substantially reduced measurement and computational costs.

\subsubsection{Restricted isometry property with Pauli observable measurements for low-rank states}
Pauli observables constitute one of the most fundamental measurement ensembles in quantum information and are widely used in practical quantum state tomography. A key theoretical property is that with high probability, random Pauli measurements satisfy the RIP, which provides a foundation for efficient recovery of low-rank quantum states.
\begin{theorem} (\cite[Theorem 2.1]{liu2011universal})
\label{thm:RIP-Pauli}
Suppose $d=2$ and let $\vrho^\star\in\C^{2^n\times 2^n}$ be a target density matrix of rank $r$. Let $\calA:\C^{2^n\times 2^n}\rightarrow\R^Q$ be the linear measurement map associated with Pauli observables, and let $0\le \delta_r<1$. When the number of coefficients satisfies
\begin{eqnarray}
\label{The requirement in K of Pauli RIP}
Q \gtrsim \frac{1}{\delta_r^2} 2^nr(\log 2^n)^6,
\end{eqnarray}
then, with high probability over matrices $\mA_1,\ldots,\mA_Q$ selected i.i.d.\ uniformly from the set of Pauli matrices $\{\mW_1,\ldots,\mW_{4^n}\}$,  $\calA$ satisfies the $(r,\delta_r)$-RIP. That is, for all rank-$r$ $\vrho$, we have
\begin{eqnarray}
    \label{The definition of Unfolding}
(1-\delta_r)\norm{\vrho}{F}^2 \le \frac{2^n}{Q}\|\calA(\vrho)\|_2^2 \le (1+\delta_r)\norm{\vrho}{F}^2,
\end{eqnarray}
where $\|\calA(\vrho)\|_2^2 = \sum_{q=1}^Q\< \mA_q, \vrho\>^2$.
\end{theorem}
\Cref{thm:RIP-Pauli} shows that random Pauli measurements approximately preserve the Frobenius norm of low-rank quantum states. In other words, the measurement energy $\frac{2^n}{Q}\|\calA(\vrho)\|_2^2$ remains close to the original state energy $\|\vrho\|_F^2$, up to a multiplicative distortion controlled by $\delta_r$. This near-isometry ensures that the essential information of a low-rank state is retained in the measurement outcomes, which forms the theoretical basis for stable recovery using only $O(2^nr(\log 2^n)^6)$ Pauli measurements.

Building on this RIP guarantee, one can further establish a quantitative recovery error bound for the constrained least-squares estimator under finite-shot measurements. Specifically, by combining the near-isometry of the measurement operator with concentration analysis of the empirical outcomes, the reconstruction error can be controlled as follows.
\begin{theorem}(\cite[Theorem 4]{qin2024optimal})
\label{upper bound recovery error of QST for Pauli measurements}
Suppose that $\vrho^\star\in\C^{2^n\times 2^n}$ is a target density matrix of rank $r$. Let $Q \geq C\cdot \frac{1}{\delta_{2r}^2} 2^nr(\log 2^n)^6$ for a positive constant $C$, so that $\calA$ satisfies $(2r,\delta_{2r})$-RIP with constant $\delta_{2r} \le 0.09$, as guaranteed by \Cref{thm:RIP-Pauli} with high probability. Supposing $M \leq O(Q/n)$, where $M$ is the number of state copies used to estimate each of the $Q$ Pauli observables, then with high probability, the solution of the constrained least-squares estimator (analogous to~\eqref{The loss function in QST for IC-POVM general}) satisfies
\begin{eqnarray}
\label{upper bound recovery error in the main paper}
\|\wh\vrho - \vrho^\star\|_F \lesssim h(\delta_{2r})\cdot \sqrt{\frac{4^nr(\log 2^n)}{QM}},
\end{eqnarray}
where
\begin{eqnarray}
\label{h-delta-function}
h(\delta_{2r}) = \frac{\sqrt{2}+\sqrt{82.55- 762.54\delta_{2r}- 843.09\delta_{2r}^2}}{1.5-15.7\delta_{2r}}.
\end{eqnarray}
\end{theorem}
\Cref{upper bound recovery error of QST for Pauli measurements} provides an explicit scaling law for the total measurement budget. In particular, to achieve a constant reconstruction accuracy of a rank-$r$ density matrix, the total number of measurement outcomes $QM$ must scale proportionally to $4^nr\log(2^n)$. This scaling is fundamentally different from the IC-POVM setting, where the sample complexity for low-rank states scales as $O(2^nr)$. The key distinction lies in how measurement information is collected. In the IC-POVM framework, each measurement outcome directly corresponds to a probability $\langle \mA_q,\vrho\rangle$, so the recovery complexity is governed directly by the intrinsic degrees of freedom of the target state. By contrast, in Pauli observable measurements, each observable $\langle \mW_q,\vrho\rangle$ in \eqref{association Pauli measurement and Pauli basis measurements} is not obtained directly as a single probability, but rather as a linear combination of $2^n$ POVM outcome probabilities. Consequently, multiple pieces of measurement information are compressed into a single observable estimate, introducing an additional statistical aggregation cost and leading to the higher scaling $O(4^nr\log(2^n))$.

Next, we discuss the trade-off between the quality and quantity of Pauli observable measurements. Recall that the RIP constant $\delta_{2r}$ decreases as the number of distinct Pauli observables $Q$ increases, while the function $h(\delta_{2r})$ is monotonically increasing in $\delta_{2r}$. Therefore, under a fixed total measurement budget $N = QM$, allocating the budget across a larger number of distinct observables (i.e., increasing $Q$ while reducing the repetition number $M$) generally leads to a tighter recovery error bound. In the extreme case of single-shot measurements ($M=1$ and $Q=N$), each empirical Pauli observable estimate is maximally noisy due to the absence of repeated sampling. Nevertheless, this strategy maximizes measurement diversity and yields the most favorable theoretical recovery guarantee. This observation provides a theoretical explanation for the effectiveness of single-sample measurements in recent scalable quantum tomography methods \cite{wang2020scalable}. A more detailed discussion of this measurement allocation principle can be found in \cite{qin2024optimal}.

\subsubsection{Recovery of low-rank states with random unitaries}

Beyond Pauli measurements, another important class of measurement schemes for low-rank quantum state tomography is based on random unitary transformations. The basic idea is to first apply a unitary transformation $\mU$ to the unknown quantum state and then perform measurements in the computational basis, thereby inducing a POVM determined by the unitary ensemble. In practice, Haar-random unitaries provide ideal measurement ensembles due to their strong randomness properties, but their implementation is computationally expensive. To address this issue, unitary $t$-designs are widely adopted as efficient approximations to Haar-random unitaries, since they reproduce the first $t$ moments of the Haar measure while remaining significantly more structured and experimentally feasible. This makes unitary $t$-design POVMs a natural framework for compressed quantum state tomography. Existing results show that the recovery performance depends critically on the order $t$ of the unitary design. In general, higher-order designs provide stronger concentration and more efficient measurement complexity, leading to improved sample complexity bounds for low-rank quantum state recovery. In particular, a key RIP-type characterization for unitary $2$-design measurements established in \cite[Eq.~(S35)]{huang2020predicting} is that, in the qubit case $d=2$, the measurement ensemble satisfies
\begin{eqnarray}
    \label{t designs_Measurement_Conclusion_Theorem}
    \E_{\calA}\|\calA(\vrho_1 - \vrho_2)\|_2^2 = \E_{\calA}\bigg[\sum_{q=1}^{Q}\sum_{k=1}^{2^n}\<\mA_{q,k}, \vrho_1 - \vrho_2  \>^2\bigg]  = \frac{Q}{2^n+1}\|\vrho_1 - \vrho_2\|_F^2.
\end{eqnarray}
Moreover, by exploiting the properties of unitary $3$-designs \cite[Eq.~(S36)]{huang2020predicting}, we further obtain
\begin{eqnarray}
    \label{unitary 3 designs_Measurement_Conclusion_Theorem}
    \E_{\calA}\bigg[\sum_{q=1}^{Q}\sum_{k=1}^{2^n}\<\mA_{q,k}, \vrho_1 - \vrho_2  \>^3\bigg] = \frac{Q}{(2^n+2)(2^n+1)}\trace\big((\vrho_1 - \vrho_2)^3\big).
\end{eqnarray}
This shows that unitary $3$-design measurements preserve higher-order moment information of the state difference, providing a more refined characterization of the recovery geometry beyond the second-order RIP structure.

Based on this characterization, we obtain the following result.

\begin{theorem}(\cite[Theorem A.1]{francca2021fast})
\label{upper bound recovery error of projected ls method}
Suppose that $\vrho^\star \in \C^{2^n \times 2^n}$ is a target density matrix of rank $r$, and let $\calA$ denote the linear measurement map induced by $Q$ POVMs sampled from a unitary $t$-design ensemble, where each POVM is measured with $M$ shots. Assume that
\begin{align}
\begin{cases}
    Q = \Omega(r^3/\epsilon^2), QM = \Omega(2^n r^3 \log 2^n / \epsilon^4), & \text{unitary 3-designs} \\
    Q = \Omega(r/\epsilon^2), QM = \Omega(2^n r \log 2^n / \epsilon^4), & \text{unitary 4-designs}.
    \end{cases}
\end{align}
Under these sampling conditions, the estimator $\wh\vrho$ produced by the Hamiltonian updates algorithm  \cite{francca2021fast}
satisfies
\begin{eqnarray}
\label{upper bound recovery error in the main paper unitary t-designs}
\|\wh\vrho - \vrho^\star\|_F \le \epsilon.
\end{eqnarray}
\end{theorem}

\Cref{upper bound recovery error of projected ls method} indicates that higher-order unitary designs significantly improve the sample complexity compared with Pauli observable measurements, reducing both the required number of measurement settings and the total number of state copies. At the same time, such gains remain weaker than those achieved by IC-POVM-based tomography, where the recovery guarantees are governed directly by the intrinsic degrees of freedom of the quantum state rather than by the design order.

\subsubsection{Stable embedding of Haar random projective measurements for MPOs and PEPOs}

The above Haar-random projective measurement frameworks can be extended to other structured quantum states, including MPO and PEPO representations. However, unlike the RIP guarantees established for random Pauli measurements in \eqref{The definition of Unfolding}, such RIP-type properties either fail to hold or hold only with significantly weaker constants for rank-one projective measurements \cite{zhong2015efficient,kueng2017low,qin2024quantum}, when moving from the exact isometry property in expectation \eqref{t designs_Measurement_Conclusion_Theorem} to finite-sample measurement operators. Consequently, standard RIP-based analyses are no longer directly applicable in these more general settings. To address this challenge, one instead establishes a weaker notion of stability, namely a one-sided lower bound on the measurement operator under Haar-random projective measurements. This stable embedding is sufficient to guarantee stable recovery for structured tensor-network states such as MPOs and PEPOs.
\begin{theorem} (\cite{qin2024quantum,qin2025quantum})
\label{Small_Method_ROHaar_Measurement}
Let $\vrho$ be an MPO  or a PEPO, and let $\calA:\C^{d^n\times d^n} \rightarrow \R^{KQ}$ be the linear mapping defined in \eqref{The defi of population measurement in Q cases (K measurements)} that is induced by $Q$ random unitary matrices. Suppose that $K\ge 1$ and
\begin{eqnarray}
    \label{number of sample rank one haar}
    Q\gtrsim \begin{cases} nd^2\ol{r}^2 \log n, & \text{MPO},\\
    nd^2\wt{r}^4 \log n, & \text{PEPO},
    \end{cases}
\end{eqnarray}
where $\bar r=\max_i r_i$ for MPO and $\widetilde r=\max_{i,j,k,l}r_{\underline{i}\,\underline{j},\underline{k}\,\underline{l}}$ for PEPO.  Then, with probability at least  $1-e^{-\alpha_1Q}$ for some constant $\alpha_1>0$, the measurement map $\calA$ satisfies
\begin{eqnarray}
    \label{Small_Method_ROHaar_Measurement_Conclusion_Theorem}
    \|\calA(\vrho)\|_2 = \bigg(\sum_{i=1}^Q\sum_{k=1}^K \<\mA_{q,k}, \vrho  \>^2 \bigg)^{\frac{1}{2}}  \gtrsim\frac{\sqrt{QK}}{d^n}\|\vrho\|_F.
\end{eqnarray}
\end{theorem}
It is worth emphasizing that the requirement on $Q$ in \eqref{number of sample rank one haar} is likely not optimal. In the proof, the randomness across different columns of each Haar-random unitary matrix is not fully exploited; instead, the analysis treats the induced projective measurements in a largely independent manner. This leads to a relatively conservative lower bound on the number of measurement settings $Q$. However, the columns of a Haar-random unitary matrix exhibit only weak local dependence, since the orthogonality constraint is a global property of the matrix \cite{tropp2012comparison}. This suggests that a sharper analysis, which more fully leverages the joint randomness structure of Haar unitary matrices, may substantially reduce the required number of measurement settings, potentially even to the minimal regime $Q=1$.

Based on the stable embedding established above, we can further derive statistical recovery guarantees for reconstructing MPOs and PEPOs using Haar random projective measurements.
\begin{theorem} (\cite{qin2024quantum,qin2025quantum})
\label{Statistical Error_Haar_Measurement}
Let $\vrho^\star$ be an MPO  or a PEPO. Suppose that $Q$ independent Haar-random unitary matrices are generated, and let $\epsilon>0$. Assume that
\begin{equation}
Q\gtrsim
\begin{cases}
nd^2\bar r^2\log n, & \textup{MPO},\\
nd^2\widetilde r^4\log n, & \textup{PEPO},
\end{cases}
\end{equation}
and
\begin{equation}
\label{Statistical Error_Haar_Measurement_Conclusion_Theorem_M}
QM\gtrsim
\begin{cases}
\frac{nd^2\bar r^2\log n(\log Q+n\log d)^2}{\epsilon^2}, & \textup{MPO},\\
\frac{nd^2\widetilde r^4\log n(\log Q+n\log d)^2}{\epsilon^2}, & \textup{PEPO},
\end{cases}
\end{equation}
where $\bar r=\max_i r_i$ for MPO and $\widetilde r=\max_{i,j,k,l}r_{\underline{i}\,\underline{j},\underline{k}\,\underline{l}}$ for PEPO. Then any global minimizer $\wh\vrho$ of the constrained least-squares estimator satisfies
\begin{equation}
\label{Statistical Error_Haar_Measurement_Conclusion_Theorem}
\|\wh{\vrho}-\vrho^\star\|_F\le \epsilon
\end{equation}
with high probability.
\end{theorem}
\Cref{Statistical Error_Haar_Measurement} shows that the total number of state copies required for reconstructing MPOs and PEPOs scales polynomially with the system size $n$, in sharp contrast to the exponential scaling required for general quantum states. This polynomial dependence reflects the low-complexity tensor-network structure and highlights the statistical advantage of exploiting such structures in quantum state tomography. Moreover, the current polynomial overhead in $n$ is mainly a consequence of the proof technique rather than an intrinsic information-theoretic limitation. In particular, the present analysis does not fully exploit the randomness structure of Haar-distributed unitary matrices. A more refined analysis may further improve the dependence on $n$ and reduce the overall sample complexity.

\subsubsection{Local measurements}
\label{sec:local-meas-result}
Compared with the global measurement schemes discussed above, local measurement settings are often easier to implement experimentally due to their reduced circuit depth and locality constraints. However, from a theoretical perspective, their analysis is significantly more challenging. In particular, RIP-type or stable embedding properties typically fail to hold in this regime, and the resulting guarantees rely instead on algorithm-specific analyses rather than geometric properties of the measurement operator.

In this section, we summarize two representative results under local measurement settings with $d=2$.

\begin{theorem}(\cite{francca2021fast})
\label{thm:local_unitary_tdesign_hamiltonian}
Suppose that $\vrho^\star \in \C^{2^n \times 2^n}$ is a target density matrix of rank $r$, and let $\calA$ be the linear measurement map induced by a local unitary $t$-design POVM. Consider the Hamiltonian updates algorithm in \cite{francca2021fast}. For local unitary $4$-designs, it suffices to sample
\[
Q = \Omega \bigg(\frac{2^{8.33n}r}{\epsilon^2} \bigg)
\]
measurement settings, with a total number of state copies
\[
QM = O\bigg(\frac{2^{13.5n} r}{ \epsilon^4}\bigg).
\]
Under these sampling conditions, the estimator $\wh\vrho$ produced by the Hamiltonian updates algorithm satisfies
\begin{equation}
\label{eq:hamiltonian_updates_error}
\|\wh\vrho - \vrho^\star\|_F \le \epsilon
\end{equation}
with high probability.
\end{theorem}

\begin{theorem}(\cite{guctua2020fast})
\label{thm:pauli_projected_ls}
Suppose that $\vrho^\star \in \C^{2^n \times 2^n}$ is a target density matrix of rank $r$, and let $\calA$ be the linear measurement map induced by Pauli basis measurements. Consider the projected least squares (PLS) estimator in \cite{guctua2020fast}. When the number of measurement settings satisfies
\[
Q = 3^n,
\quad
QM =O\bigg(\frac{3^{n} r}{ \epsilon^2}\bigg),
\]
the estimator $\wh\vrho$ produced by the PLS method satisfies
\begin{equation}
\label{eq:pls_error}
\|\wh\vrho - \vrho^\star\|_F \le \epsilon
\end{equation}
with high probability.
\end{theorem}
These results highlight a fundamental distinction between local and global measurement frameworks. In contrast to global schemes where geometric properties such as RIP or stable embeddings can be established, local measurement settings do not admit such unified structures. As a consequence, their theoretical guarantees rely heavily on algorithm-specific analyses and typically exhibit significantly higher sample complexity than the intrinsic degrees of freedom of low-rank states, which scale as $2^n r$. This reflects the intrinsic difficulty of extracting global information from strictly local measurements.

\subsection{Optimization methods}
In this section, we review optimization methods for estimating the underlying quantum state from measurements. While some of the statistical guarantees reviewed in previous sections are derived for specific algorithms, our goal here is to provide a broader perspective on optimization-based approaches for QST.

For general quantum states, a variety of methods have been developed for QST. Classical approaches include linear inversion \cite{fano1957description}, maximum likelihood estimation (MLE) \cite{vrehavcek2001iterative,fiuravsek2001maximum,lvovsky2004iterative,haffner2005scalable,shang2017superfast,bolduc2017projected}, Bayesian inference \cite{blume2010optimal,granade2016practical,lukens2020practical}, confidence or region estimation \cite{blume2012robust,faist2016practical}, least-squares estimators \cite{kyrillidis2018provable,brandao2020fast,zhu2024connection}, approximate message assing~\cite{siekierski2025approximate}, and more recently classical machine learning approaches \cite{lohani2020machine}. Beyond these classical formulations, QST has also been studied through  variational frameworks, including quantum machine learning (QML) methods such as variational quantum circuits, hybrid quantum-classical optimization, and related quantum-enhanced estimation approaches \cite{sen2022variational,liu2020variational,kurmapu2020machine}. These methods typically parameterize the quantum state or measurement process and optimize the parameters using measurement feedback. A broader overview of these formulations can be found in \cite{innan2024quantum}.

In this survey, we focus primarily on optimization principles that arise across different classical QST formulations, particularly constrained optimization, projected methods, and parameterized nonconvex optimization for structured quantum states. Given quantum measurements $\wh\vp$ of an unknown physical state $\vrho^\star \in \setX$, we seek to recover the state by solving the constrained optimization problem
\begin{align}
\wh \vrho
=
\argmin_{\vrho\in\setX}
\ell(\calA(\vrho),\wh\vp),
\label{eq:problem-qst general}
\end{align}
where $\ell$ is a loss function that measures the discrepancy between the predicted measurements $\calA(\vrho)$ and the observed measurements $\wh\vp$. Common choices include the least-squares loss \cite{kyrillidis2018provable,brandao2020fast} and the negative log-likelihood loss \cite{hradil1997quantum,vrehavcek2001iterative}.

For many structured quantum states described in \Cref{sec:structured-states}, the feasible set $\setX$ is nonconvex, which makes \eqref{eq:problem-qst general} a nonconvex optimization problem. For certain structures, such as low-rank states, convex relaxation approaches (e.g., nuclear norm minimization) have been successfully developed and can provide strong statistical recovery guarantees under suitable measurement assumptions \cite{gross2010quantum,flammia2012quantum}. However, such convex relaxations are not always available or tight for more general structured state classes, such as MPS, tensor network states, or neural quantum states. In these settings, the structure is typically encoded through highly nonlinear parameterizations, making it challenging to construct tractable convex surrogates that faithfully preserve the original geometry or sample complexity advantages.

Fortunately, a broad class of iterative optimization methods, particularly gradient-based approaches, have been developed for such nonconvex formulations and have demonstrated strong empirical and theoretical performance. More broadly, a large body of work in nonconvex optimization has shown that, despite the lack of global convexity, iterative algorithms can often converge to globally optimal or statistically optimal solutions under suitable initialization, regularity, or measurement conditions \cite{jain2017non,chi2019nonconvex,zhang2020symmetry}. This perspective underlies many optimization-based approaches for structured QST.

\paragraph{Projected gradient descent.} To incorporate the structural constraint $\setX$, a natural approach is projected gradient descent, also commonly referred to as iterative hard thresholding (IHT) in the compressive sensing and low-rank recovery literature \cite{gross2010quantum,lidiak2022quantum,shanmugam2023unrolling,qin2024quantum}. At each iteration, the method performs a gradient-based update followed by a projection onto the structured set, defined as
    \begin{eqnarray}
    \label{equation of IHT}
    \vrho^{(t+1)} = \calP_{\setX}(\vrho^{(t)} - \mu \nabla_{\vrho} \ell(\calA(\vrho^{(t)}),\wh\vp) ),
    \end{eqnarray}
    where $\calP_{\setX}$ denotes the optimal projection onto the target set $\setX$, i.e., $\calP_{\setX}(\vrho') := \argmin_{\vrho \in \setX}\norm{\vrho - \vrho'}{F}$, and $\mu$ is a step size. When $\setX$ admits a smooth manifold structure and tangent-space operations are efficient to compute, the Euclidean gradient can also be replaced by a Riemannian gradient, leading to manifold-based optimization methods \cite{hsu2024quantum}.

The computational complexity of $\calP_{\setX}$ depends critically on the structure of $\setX$. For low-rank states, the projection can be computed efficiently via eigendecomposition or singular value truncation, making projected gradient descent attractive in low-rank QST \cite{hsu2024quantum}. However, for more general structured state classes, computing the exact projection $\calP_{\setX}$ can be challenging and, in some cases, computationally intractable. For example, for MPS or more generally MPO representations, no exact tractable algorithm is known for computing the globally optimal projection. In practice, approximate projection methods such as TT-SVD \cite{oseledets2011tensor} are often employed to compute low-rank tensor approximations with suboptimal guarantees, and have been observed to perform well for QST  \cite{cramer2010efficient,qin2024sample}.  The projection problem becomes even more challenging for higher-dimensional tensor-network states such as PEPOs. Unlike MPO or MPS representations, PEPO networks contain loops in the underlying tensor-network geometry, which prevents efficient sequential truncation procedures such as TT-SVD. While several approximate compression or variational truncation methods have been proposed, no tractable general-purpose algorithm is currently known for computing the exact globally optimal projection onto the PEPO set while simultaneously enforcing physical constraints such as positive semidefiniteness \cite{orus2019tensor}. Similar difficulties also arise for neural quantum states, where exact projection onto the structured state manifold is generally computationally challenging due to the highly nonlinear parameterization.

To address this limitation, a broader and often more scalable alternative is to directly optimize over a parameterized representation of the structured state, thereby avoiding explicit projection onto $\setX$.

\paragraph{Parameterized approach.}
As discussed in \Cref{sec:unified-notation-structure}, many structured quantum states admit explicit parameterizations of the form $
\setX = \{\vrho_{\vtheta} : \vtheta \in \Theta\}.$ In this setting, rather than optimizing directly over the state $\vrho$, one can optimize over the parameter space $\Theta$:
\begin{align}
\min_{\vtheta \in \Theta}
\ell\!\bigl(\calA(\vrho_{\vtheta}), \wh\vp\bigr).
\label{eq:problem-qst-parameterized}
\end{align}
Compared with the original constrained optimization problem \eqref{eq:problem-qst general}, the parameterized formulation \eqref{eq:problem-qst-parameterized} incorporates the structural constraint directly into the model parameterization. As a result, the optimization is performed in a lower-dimensional and often more structured parameter space.

To solve \eqref{eq:problem-qst-parameterized}, a similar projected gradient descent can be applied to \eqref{eq:problem-qst-parameterized}:
\e
\vtheta^{(t+1)}
=
\calP_{\Theta}
\!\left(
\vtheta^{(t)}
-
\mu
\nabla_{\vtheta}
\ell\!\bigl(\calA(\vrho_{\vtheta^{(t)}}),\wh\vp\bigr)
\right),
\label{eq:PGD-parameterized}\ee
where $\calP_{\Theta}$ denotes the projection onto the feasible parameter set $\Theta$. Compared with the projection step in \eqref{equation of IHT}, projection in the parameterized formulation is often significantly easier to compute. For example, in low-rank factorized formulations, the projection often reduces to normalization onto the unit Frobenius sphere. For MPS or MPO parameterizations, the projection is also relatively simple depending on the chosen gauge or canonical form. In neural quantum states, $\Theta$ is often taken as the full parameter space, in which case the projection operator reduces to the identity map. More generally, this parameterized viewpoint transforms the original constrained optimization over density matrices into a structured nonconvex optimization problem over model parameters, which is often substantially more scalable in practice.

On the other hand, the parameterized problem \eqref{eq:problem-qst-parameterized} is often nonconvex, so local search methods such as \eqref{eq:PGD-parameterized} do not, in general, guarantee convergence to a globally optimal solution. More broadly, nonconvex optimization can exhibit spurious local minima, saddle points, and poor conditioning, and even finding critical points may be computationally intractable in the worst case \cite{murty1987some}. Fortunately, for several important structured state classes, the resulting optimization problem often exhibits favorable geometric properties that enable efficient optimization by gradient-based methods. For low-rank quantum states, this formulation is closely related to the \emph{Burer--Monteiro factorization} \cite{burer2003nonlinear,burer2005local}, which parameterizes a positive semidefinite low-rank matrix through a factorized representation. This perspective has been extensively studied in signal processing, machine learning, and optimization \cite{bhojanapalli2016global,ge2016matrix,zhu2018global}; see \cite{chi2019nonconvex} for a comprehensive review. Although the resulting problem is nonconvex, under suitable regularity and measurement conditions it often admits a benign optimization landscape, enabling gradient-based methods to converge to globally optimal or statistically optimal solutions. In the context of QST, such factorized formulations have been exploited for low-rank recovery, with convergence analysis established in \cite{kyrillidis2018provable} and more recent landscape analysis in \cite{qin2026optimal}. Convergence guarantees for MPS and MPO in general inverse problem settings have also been studied in \cite{cai2022provable,qin2024guaranteed,qin2025enhancing}.

For neural quantum states, parameterized optimization is the default training paradigm, where the quantum state is represented by a neural network and optimized directly through gradient-based methods. However, compared with low-rank and tensor-network settings, formal convergence guarantees remain much less understood due to the highly nonlinear, overparameterized, and strongly nonconvex nature of the optimization landscape.

\subsection{Projected classical shadow estimation}
In this section, we review classical shadow methods, which can be viewed as approximate approaches for solving \eqref{eq:problem-qst general} under randomized POVMs. Compared with the iterative optimization methods discussed above, classical shadow methods provide a non-iterative framework for efficiently estimating quantum observables from measurement data, resulting in computationally efficient post-processing. Moreover, these estimators can be naturally combined with structured recovery procedures, serving as effective initializations for iterative optimization-based reconstruction methods, as discussed later.

\subsubsection{Classical shadows}
To motivate the construction of classical shadows, consider the least-squares formulation of \eqref{eq:problem-qst general} while temporarily ignoring the physical constraints on the density matrix:
\begin{align*}
	\wh \vrho = \argmin_{\vrho'\in \C^{d^n\times d^n}} ~  \|\calA(\vrho') - \wh\vp\|_2^2.
    \end{align*}
The corresponding solution $\wh \vrho$ satisfies the normal equation
\e
\calA^\dagger\calA(\wh\vrho) = \calA^\dagger(\wh \vp),
\ee
where $\calA^\dagger$ denotes the adjoint of $\calA$.
When the ensemble of $Q$ POVMs is informationally complete, $\calA^\dagger \calA$ is invertible and the solution is unique, given by $\wh \vrho = \parans{\calA^\dagger\calA}^{-1} \parans{\calA^\dagger(\wh \vp)}$. In contrast, when the POVMs are not informationally complete, which is often the case in practical compressive measurement settings, the operator $\calA^\dagger\calA$ becomes rank-deficient and the least-squares problem admits infinitely many solutions. Standard approaches in this setting include adding $\ell_2$ regularization or using the pseudoinverse of $\calA^\dagger\calA$ to select the minimum-norm solution~\cite{zhu2024connection}.

Classical shadow estimation adopts a different strategy for stabilizing the inverse problem by exploiting the statistical properties of the randomized POVM ensemble. Specifically, suppose each POVM $\{\mA_{q,k}\}_{k\in[K]}$ is independently and randomly generated from an ensemble of POVMs $\setA$ according to a certain probability distribution $P(\setA)$. As the classical shadow aims to significantly reduce experimental cost, the regime of interest is the case that the linear map $\calA$ in \eqref{The defi of population measurement in Q cases (K measurements)} for obtaining quantum measurements is not informationally complete. In this case, we may approximate $\frac{1}{Q}\calA^\dagger \calA$ by its expectation
\begin{align}
\begin{split}
\calM(\vrho)  =\E\left[\frac{1}{Q}\calA^\dagger \calA(\vrho)\right]  = \E_{\{\mA_{k}\}\sim P(\setA) } \left[\sum_{k=1}^K  \innerprod{\mA_{k}}{\vrho} \mA_{k} \right],
\label{eq:Exp-AtA}
\end{split}
\end{align}
where $\{\mA_{k}\}_{k\in[K]}$ represents a random POVM generated from the ensemble of POVMs $\setA$ according to the probability distribution $P(\setA)$. Here $\calM$ is called the quantum channel.  If the ensemble $\setA$ is informationally complete in expectation, then $\calM$ is invertible. The classical shadow estimation  introduced in \cite{huang2020predicting} is then given by
\begin{align}
\vrho_{\textup{shadow}} = \calM^{-1}(\calA^\dagger(\wh \vp)) = \frac{1}{Q} \sum_{q=1}^Q \underbrace{\calM^{-1}\left(\calA_q^\dagger(\wh \vp_q)\right)}_{\text{classical shadow}},
\label{eq:shadow-v3}
\end{align}
where $\wh\vrho_q \vcentcolon= \mathcal{S}\left(\calA_q^\dagger(\wh \vp_q)\right) = \calM^{-1}\left(\calA_q^\dagger(\wh \vp_q)\right)$ is called a classical shadow. As $\calM$ defined in \eqref{eq:Exp-AtA} is a linear operator, its inverse $\calM^{-1}$ is also linear.  These classical shadows are independent and unbiased estimators of $\vrho$.

The form of the quantum channel $\calM$ depends on the measurement ensemble. A widely studied setting considers rank-one projective measurements induced by random unitary matrices; that is rank-1 POVMs of form $\{\mA_{1},\ldots,\mA_{d^n} \}$ with $\mA_{q,k_q} = \vu_{q,k_q}\vu_{q,k_q}^\dagger$, where each $\mU_q = \begin{bmatrix} \vu_{1} & \cdots & \vu_{d^n} \end{bmatrix}$ is randomly chosen from an ensemble of $d^n\times  d^n$ unitary matrices $\setU$.  Various unitary ensembles have been explored in prior studies, including the local and global Clifford ensembles \cite{huang2020predicting}, fermionic Gaussian unitaries \cite{zhao2021fermionic}, chaotic Hamiltonian evolutions \cite{hu2023classical}, locally scrambled unitary ensembles \cite{hu2023classical}, and Pauli-invariant unitary ensembles \cite{bu2022classical}, for which explicit formulas for the quantum channel $\calM$ and its inverse $\calM^{-1}$ exist. For example, when $\setU$ is the full unitary group equipped with the Haar measure, the corresponding quantum channel and its inverse admit explicit formulas:
\begin{align}
 \calM(\vrho)  = \frac{1}{d^n+1}\vrho + \frac{\trace(\vrho)}{d^n+1}\mId, \quad
 \calM^{-1}(\vrho) = (d^n + 1)\vrho - \trace(\vrho)\mId.
\label{eq:shadow-global-haar}\end{align}
Substituting these expressions into \eqref{eq:shadow-v3} yields the classical shadow estimator
$
 \wh \vrho_q =(d^n + 1) (\mU_q \wh \vp_q)(\mU_q \wh \vp_q)^\dagger - \mId.
$
Since $(\mU_q \wh \vp_q)(\mU_q \wh \vp_q)^\dagger$ is rank-1 with its only non-zero eigenvalue equal to 1, each classical shadow satisfies $\trace(\wh\vrho_q)=1$, although it is generally not positive semidefinite; one positive eigenvalue is equal to $d^n$ and $d^n-1$ negative eigenvalues are equal to $-1$.

\paragraph{Predicting quantum properties.}
In many applications, one does not require reconstruction of the entire density matrix. Instead, the goal is to estimate specific properties of the quantum state, such as fidelities, local observables, entanglement measures, or correlation functions. A large body of prior work has developed specialized approaches for such tasks, including direct fidelity estimation \cite{da2011practical,flammia2011direct,seshadri2024theory}, local observable estimation \cite{bonet2020nearly}, entanglement estimation \cite{horodecki2002method,horodecki2003measuring,mintert2007observable}, and classical shadow methods \cite{huang2020predicting}.  A comprehensive review of quantum property estimation is beyond the scope of this survey. We therefore only briefly mention these directions here and refer readers to related references therein.

Among these approaches, many are tailored to particular observables and require task-specific measurements or postprocessing. Classical shadows, by contrast, provide a more unified framework. Instead of designing measurements separately for each target property, classical shadows first collect randomized measurements from a fixed ensemble and then reuse the resulting classical data to estimate many observables simultaneously. This decouples the measurement process from downstream prediction tasks and enables efficient post hoc estimation of a large collection of quantum properties.

Consider estimating a linear observable $\trace(\mB\vrho^\star)$ using classical shadows $\{\wh \vrho_q\}_q$. Since each classical shadow $\wh \vrho_q$ is an independent and unbiased estimator of $\vrho^\star$, the quantity $\trace(\mB\wh \vrho_q)$ is also an independent and unbiased estimator of $\trace(\mB\vrho^\star)$. Therefore, the collection $\{\trace(\mB\wh \vrho_q)\}_q$ can be used to estimate $\trace(\mB\vrho^\star)$. The estimation accuracy can be further improved using robust estimation procedures, such as the median-of-means estimator, which provide stronger sample complexity guarantees. The following result summarizes the corresponding performance guarantee.
\begin{theorem} (\cite[Theorem 1]{huang2020predicting})
Classical shadows of size $Q$ suffice to predict $N$ arbitrary linear observables
\begin{eqnarray}
\label{set of the target functions}
\trace(\mB_1\vrho^\star),\ldots,\trace(\mB_N\vrho^\star)
\end{eqnarray}
up to additive error $|\trace(\mB_i\vrho_{\textup{CS}}) - \trace(\mB_i\vrho^\star)|\leq\varepsilon,\ i=1,\dots,N$, provided that
\begin{eqnarray}
\label{the requirement of sample complexity}
Q \ge O\!\left(
\log(N)\max_{i}\frac{\|\mB_i\|_{\rm shadow}^2}{\varepsilon^2}
\right).
\end{eqnarray}
The definition of the norm $\|\mB_i\|_{\rm shadow}$ depends on the ensemble of unitary transformations used to construct the classical shadow.
\label{thm:CS-est-property}\end{theorem}
The above result highlights the fundamental property of classical shadows: a logarithmic dependence on the number of target observables. In particular, although the sample complexity $Q$ depends only logarithmically on $N$, it scales linearly with the shadow norm $\|\mB_i\|_{\rm shadow}^2$, which captures the interaction between the measurement ensemble and the target observables.
In our setting, this implies that $N$ should be interpreted as the number of quantum properties of interest like fidelity, entanglement measures, and correlations. In other words, classical shadows enable the simultaneous prediction of a large collection of observables from a relatively small number of state copies, with the complexity governed primarily by the intrinsic difficulty of the observables.
This perspective is particularly important when the goal is not full state reconstruction but efficient estimation of structured or task-oriented properties of $\vrho^\star$.

\subsubsection{Projected classical shadows for quantum state tomography}
The discussion above focuses on estimating quantum properties of interest, where the goal is to efficiently predict expectation values of a collection of observables. We now turn to reconstruction-based methods under the same randomized measurement model, where the objective is to recover the underlying quantum state from the measurement outcomes. The key idea is to exploit structural information about the underlying state by projecting the classical shadow estimator onto the structured set $\setX$, referred to as the projected classical shadow estimator
\begin{eqnarray}
    \label{projected classical shadow any set}
    \vrho_{\text{proj-shadow}} =  \calP_{\setX}(\vrho_{\text{shadow}}) := \argmin_{\vrho \in \setX}\norm{\vrho - \vrho_{\text{shadow}}}{F}.
\end{eqnarray}
Intuitively, this projection step incorporates prior structural information and converts a generic unbiased estimator into a structure-aware reconstruction procedure. The following result establishes guarantees.
\begin{theorem} (\cite{qin2025enhancing})
\label{project CS:general}
For a target state $\vrho^\star\in\setX$, consider global Haar-random projective measurements. Assume
\[
M\gtrsim \frac{\log \calN(\setX)}{\epsilon^2}
\]
for any given $\epsilon>0$, where $\calN(\setX)$ is the covering number descripbed in \Cref{sec:unified-notation-structure}.
Then, with high probability, the projected classical shadow $\vrho_{\textup{proj-shadow}}$ in \eqref{projected classical shadow any set} with the classical shadow estimator  defined in \eqref{eq:shadow-v3} and \eqref{eq:shadow-global-haar} satisfies
\begin{equation}
\label{error bound of projected classical shadow MPO F1}
\|\vrho_{\textup{proj-shadow}}-\vrho^\star\|_F\le \epsilon.
\end{equation}
\end{theorem}
\Cref{project CS:general} shows that the projected classical shadow estimator achieves information-theoretically optimal recovery rates for a broad class of structured quantum states, provided that the projection onto $\setX$ can be computed exactly or sufficiently well approximated. As discussed in the previous section, for low-rank states, this projection can be computed efficiently. For more general structured state classes such as MPS, computing the exact globally optimal projection is typically intractable. However, practical approximate projection methods, such as tensor truncation or TT-SVD-based procedures, often provide high-quality suboptimal projections. Accordingly, the recovery guarantees can be extended to approximate projection settings, typically with an additional approximation error term reflecting the projection inaccuracy \cite{qin2025enhancing}.

\paragraph{Connection with spectral initialization and projected least squares.}
Projected classical shadows closely resembles spectral initialization methods used in nonconvex optimization \cite{lu2020phase}. Specifically, in our setting, spectral initialization produces the estimate $\calP_{\setX}(\calA^\dagger(\wh{\vp}))$. Rewriting the projected classical shadow estimator in \eqref{projected classical shadow any set} as $
\calP_{\setX}\!\left(\calM^{-1}\!\left(\calA^\dagger(\wh{\vp})\right)\right),$
it can be interpreted as a preconditioned spectral initialization method, where the inverse quantum channel $\calM^{-1}$ serves as a preconditioner. This preconditioning corrects for the bias induced by the measurement ensemble and improves recovery under random measurements. Projected classical shadow estimation is also closely related to, and can be viewed as a generalization of, projected least squares \cite{guctua2020fast}. Specifically, when $\calA$ is deterministic and informationally complete, then $\calM = \calA^\dagger \calA$, $\vrho_{\text{shadow}}$ reduces to the least-squares estimator, and the projected classical shadow becomes exactly the projected least-squares estimator. From this perspective, projected classical shadows extends projected least squares to the setting of random measurements. Finally, we note that while projected classical shadows retains strong statistical guarantees, as with spectral initialization, it can also be further refined using iterative algorithms described in the previous section.

\section{Discussion and Conclusion}
\label{sec:discussion-conclusion}

In this paper, we review recent progress on QST for structured quantum states, with an emphasis on the interplay between sample complexity, measurement design, and optimization. A central message is that compact representations are fundamental to both statistical and algorithmic efficiency. For general quantum states, tomography requires resources that scale exponentially with the number of qudits. In contrast, structured state classes such as low-rank states, MPOs, PEPOs, shallow-circuit states, and neural quantum states can reduce the effective degrees of freedom and thereby enable substantially improved sample complexity.

This highlights the importance of developing new representation families for large-scale quantum systems. In particular, neural quantum representations have emerged as a promising direction due to their expressive power and flexibility \cite{lange2024architectures}. However, their statistical complexity, optimization landscape, and approximation-theoretic properties remain significantly less understood than those of low-rank and tensor-network models. For example, deriving tight geometric complexity bounds for low-rank states and MPOs in \eqref{covering number of different structures} relies heavily on canonical or structured parameterizations, which are often unavailable or less explicit for neural-network-based representations. On the other hand, simply counting the number of parameters is typically insufficient for characterizing model complexity, due to the strong nonlinearity, compositional structure, and frequent over-parameterization of neural networks. As a result, obtaining sharp sample-complexity bounds, analyzing the resulting optimization landscape, and establishing rigorous recovery guarantees remain significant challenges for neural QST.

Another important direction concerns local measurements. Local measurement schemes are experimentally attractive because they are easier to implement on near-term quantum devices. However, as discussed in \Cref{sec:local-meas-result}, their theoretical analysis is considerably more challenging than that of global measurement schemes. One difficulty is that the elements of local POVMs involve products of local operators, as in \eqref{elements in a set of local POVM}, which creates strong structural dependence and makes standard concentration arguments much harder to apply. As a result, existing guarantees for local measurements are often algorithm-specific and do not yet provide a unified stable-embedding or RIP-type theory comparable to that available for global measurements. Developing new statistical tools for local measurement ensembles is therefore an important open problem. In addition, recent results further suggest that the power of local measurements may depend delicately on the target state family. For example, \cite{jameson2024optimal}
shows that general MPS/MPO states can be recovered with bounded error from local measurements using only a number of state copies polynomial in the number of qubits, whereas certain long-range entangled states, such as a family of generalized GHZ states, cannot be recovered with the same efficiency. This raises a fundamental question: do local measurements provide uniform recovery guarantees over broad structured state classes, as global measurements often do, or are such guarantees restricted to specific subclasses of structured states with favorable locality or entanglement properties?

A third important issue is convergence guarantees for optimization-based reconstruction. Many practical algorithms for structured QST rely on nonconvex formulations, including low-rank factorization, tensor-network parameterizations, and neural quantum states. For low-rank models and certain tensor-network inverse problems, recent work has begun to establish benign landscape properties and convergence guarantees. However, a general theory remains incomplete, especially for higher-dimensional tensor networks such as PEPS/PEPO for lattice systems, and neural quantum states. Understanding when computationally efficient algorithms can attain the statistically optimal rates predicted by sample complexity theory remains a major challenge.

Beyond these limitations within the current structured QST framework, several broader extensions remain promising directions for future work. First, this survey primarily focuses on reconstructing the full quantum state, while only briefly touching on the related problem of estimating physically meaningful properties of complex quantum systems, such as multipartite entanglement, topological order, many-body localization, and emergent quantum phases, without reconstructing the full density matrix \cite{aaronson2019shadow}. Classical-shadow-based methods provide a resource-efficient framework for estimating many such properties from measurement data, with guarantees such as those summarized in \Cref{thm:CS-est-property}. However, the associated shadow norm can scale unfavorably with the complexity of the target observable, which may in turn require exponentially many samples. Incorporating structural priors into property estimation is therefore an important and promising direction.

Second, this survey has focused primarily on nonadaptive measurement designs. Adaptive measurements have been widely studied in the literature \cite{berry2002adaptive,quek2021adaptive} and provide a promising direction for further improving the efficiency of structured QST. The benefit of adaptivity depends on the state family, measurement model, and performance criterion. For general states, \cite{chen2023does} showed that adaptivity does not improve the minimax sample complexity for trace-distance learning, although it can improve guarantees under infidelity. In contrast, recent work \cite{goldar2026exponential} demonstrates that for a structured family of multi-qubit states, adaptive strategies achieve polynomial sample complexity, whereas every nonadaptive design requires exponentially many state copies. These results suggest that adaptivity may be particularly powerful for certain structured state classes. A systematic understanding of when and why adaptivity improves sample complexity for general structured quantum states remains an important direction for future work.

Third, this survey primarily considers statistical error arising from finite-shot measurements, while largely neglecting more realistic sources of noise such as measurement noise, device imperfections, and adversarial perturbations. In practice, robustness to such noise is essential for reliable large-scale QST \cite{cai2023quantum}. Recent work \cite{aliakbarpour2025adversarially} suggests that exploiting low-dimensional structure, such as low-rankness, can enable recovery algorithms that remain robust under adversarial corruption. More broadly, understanding how structural priors can improve robustness, error mitigation, and stability under realistic noise models represents an important direction.

\section*{Acknowledgments}
\label{sec: ack}

We acknowledge funding support from NSF Grants No. CCF-2106834, CCF-2241298
and ECCS-2409701. We are grateful to Zhexuan Gong, Alireza Goldar, and Stephen Becker for many valuable discussions.


\end{document}

%% file: macro.tex
\usepackage{url}
\usepackage[hidelinks]{hyperref}
\usepackage{amsmath,amsthm,amssymb,amsbsy}
\usepackage{paralist}
\usepackage{xcolor}
\usepackage{color}
\usepackage{graphicx}
\graphicspath{{./figs/}}
\usepackage{algorithm}
\usepackage{algorithmic}
\usepackage{comment}
\usepackage{multirow}
\usepackage{braket}
\usepackage{mathtools}

\usepackage{graphicx}
\usepackage{fancyhdr}
\usepackage{cite}
\usepackage{cleveref}
\usepackage{subcaption}
\usepackage{tcolorbox}

\newtheorem{lemma}{Lemma}
\newtheorem{defi}{Definition}

\newtheorem{theorem}{Theorem}

\newcommand{\R}{\mathbb{R}}

\newcommand{\C}{\mathbb{C}}

\newcommand{\e}{\begin{equation}}
\newcommand{\ee}{\end{equation}}
\newcommand{\en}{\begin{equation*}}
\newcommand{\een}{\end{equation*}}
\newcommand{\eqn}{\begin{eqnarray}}
\newcommand{\eeqn}{\end{eqnarray}}
\newcommand{\bmat}{\begin{bmatrix}}
\newcommand{\emat}{\end{bmatrix}}

\DeclareMathAlphabet\mathbfcal{OMS}{cmsy}{b}{n}

\newcommand{\E}{\operatorname{\mathbb{E}}}

\newcommand{\vct}[1]{\boldsymbol{#1}}
\newcommand{\mtx}[1]{\boldsymbol{#1}}

\newcommand{\<}{\langle}
\renewcommand{\>}{\rangle}

\newcommand{\trace}{\operatorname{trace}}

\newcommand{\diag}{\operatorname{diag}}

\newcommand{\nqbit}{n}
\renewcommand{\set}[1]{\mathbb{#1}}

\DeclareMathOperator*{\argmin}{\text{arg~min}}

\newcommand{\wh}{\widehat}

\newcommand{\wt}{\widetilde}
\newcommand{\ol}{\overline}
\newcommand{\ul}{\underline}

\newcommand{\norm}[2]{\left\| #1 \right\|_{#2}}

\newcommand{\parans}[1]{\left(#1\right)}
\newcommand{\innerprod}[2]{\left\langle #1,  #2 \right\rangle}

\newcommand{\calA}{\mathcal{A}}

\newcommand{\calE}{\mathcal{E}}
\newcommand{\calF}{\mathcal{F}}

\newcommand{\calM}{\mathcal{M}}
\newcommand{\calN}{\mathcal{N}}

\newcommand{\calP}{\mathcal{P}}

\newcommand{\va}{\vct{a}}

\newcommand{\ve}{\vct{e}}
\newcommand{\vf}{\vct{f}}

\newcommand{\vn}{\vct{n}}

\newcommand{\vp}{\vct{p}}

\newcommand{\vr}{\vct{r}}

\newcommand{\vu}{\vct{u}}

\newcommand{\vw}{\vct{w}}
\newcommand{\vx}{\vct{x}}
\newcommand{\vy}{\vct{y}}

\newcommand{\vtheta}{\vct{\theta}}

\newcommand{\vphi}{\vct{\phi}}
\newcommand{\vpsi}{\vct{\psi}}

\newcommand{\vrho}{\vct{\rho}}
\newcommand{\vzero}{\vct{0}}

\newcommand{\mA}{\mtx{A}}
\newcommand{\mB}{\mtx{B}}

\newcommand{\mE}{\mtx{E}}

\newcommand{\mU}{\mtx{U}}

\newcommand{\mW}{\mtx{W}}
\newcommand{\mX}{\mtx{X}}

\newcommand{\mSigma}{\mtx{\Sigma}}

\newcommand{\mId}{{\bf I}}

\newcommand{\setA}{\set{A}}

\newcommand{\setU}{\set{U}}

\newcommand{\setX}{\set{X}}

\graphicspath{{./figs/}}

\newlength{\imgwidth}
\newboolean{twoColVersion}
\setboolean{twoColVersion}{false}
\newcommand{\twoCol}[2]{\ifthenelse{\boolean{twoColVersion}} {#1} {#2} }